\documentclass[a4paper,11pt]{article}
\pdfoutput=1 

\usepackage{jcappub} 

\usepackage[T1]{fontenc} 

\usepackage{natbib}
\title{\boldmath Chemical Signatures of Population III Stars in Damped Lyman-$\alpha$ Absorption Systems at $z \approx 6$}

\author[a]{Eli Visbal,}
\author[b]{Greg L. Bryan,}
\author[b,c,d]{Zolt\'an Haiman}

\affiliation[a]{University of Toledo,
Department of Physics and Astronomy and Ritter Astrophysical Research Center, 2801 W. Bancroft Street,
Toledo, Ohio 43606. USA}

\affiliation[b]{Department of Astronomy, Columbia University, 550 West 120th Street, New York, NY, 10027, USA}

\affiliation[c]{Department of Physics, Columbia University, 550 West 120th Street, New York, NY, 10027, USA}

\affiliation[d]{Institute of Science and Technology Austria (ISTA), Am Campus 1, Klosterneuburg, Austria}

\emailAdd{Elijah.Visbal@utoledo.edu}

\abstract{Recently, Sodini et al. (2024) presented a sample of OI damped Lyman-$\alpha$ absorption system (DLA) analogs at $z\sim6$ that contain possible chemical signatures of Population III (Pop III) stars. In this paper, we use an N-body simulation-based semi-analytic model of the first stars and galaxies to predict the impact of Pop III stars on high-redshift DLAs. These Pop III DLA predictions are the first to include a number of important physical effects such as Lyman-Werner (LW) feedback, reionization, and external metal enrichment (all of which account for three-dimensional spatial fluctuations caused by halo clustering). We predict the abundance of DLAs as a function of their carbon-to-oxygen ratios ([C/O]). We find that our fiducial model is strongly ruled out by the data as it contains too few high-[C/O] DLAs, which have metals primarily from Pop III stars. However, increasing the delay time between Pop III and metal-enriched star formation due to supernovae feedback leads to better agreement with the data. Our results suggest that DLA analogs at $z\sim6$ are a promising probe of Pop III star formation for two key reasons. First, for reasonable parameter choices there are significant numbers of DLAs with metals primarily originating from Pop III stars. Second, we find that the number of DLAs with substantial Pop III contributions depends strongly on the Pop III star formation efficiency and the delay time between Pop III and metal-enriched star formation.
}

\begin{document}
\maketitle
\flushbottom

\section{Introduction}
Understanding the formation of the first stars and galaxies is currently an exciting frontier in astrophysics and cosmology. Simulations predict that the first stars formed via molecular hydrogen cooling in low-mass dark matter ``minihalos'' with virial masses of $M_{\rm h}\approx 10^{5-6} ~ M_\odot$  \cite{1996ApJ...464..523H, 1999ApJ...527L...5B, 2002Sci...295...93A, 2003ApJ...592..645Y, 2009Sci...325..601T, 2010MNRAS.403...45S, 2015ComAC...2....3G, 2023ARA&A..61...65K}. These so-called Population III (Pop III) stars are defined by their zero or extremely low metallicity and are predicted to have higher masses compared to metal-enriched stars \citep[e.g.,][]{2015MNRAS.448..568H}. Although Pop III star formation is only expected to be the dominant mode of star formation at very high redshifts \citep[e.g.,][]{2024ApJ...962...62F}, some Pop III star formation in low-metallicity regions is predicted to persist to much later times ($z\approx 6$) as a consequence of the highly inhomogeneous process of metal enrichment throughout the universe (see Fig.~1 of \cite{2023ARA&A..61...65K} or \cite{2006Natur.440..501J}).

Because they generally form in low-mass minihalos at high redshifts, Pop III stars are challenging to observe directly. There are currently no unambiguous direct directions of Pop III stars. However, we note that some high-redshift galaxies recently discovered with the \emph{James Webb Space Telescope} (\emph{JWST}) contain possible Pop III signatures (e.g., strong HeII 1640 \AA ~emission \cite{2000ApJ...528L..65T,2001ApJ...553...73O} and low metallicity) \cite{2023A&A...678A.173V,2024A&A...687A..67M,2024ApJ...967L..42W, 2025arXiv250111678F}. Direct observations of Pop III stars may require next-generation telescopes \cite{2020ApJ...904..145S} or strong gravitational lensing with caustics \cite{2018ApJS..234...41W, schauer2022probability}. Although these direct measurements may remain challenging in the near future, there are a variety of other observables that provide (or will soon provide) information on the first stars. These include the optical depth of the cosmic microwave background \cite{2015MNRAS.453.4456V}, the cosmological 21cm signal \cite{2018MNRAS.478.5591M,2022MNRAS.516..841G}, HeII 1640 line-intensity mapping \cite{2015MNRAS.450.2506V, 2022ApJ...933..141P}, supernovae or gamma ray bursts from Pop III progenitors \cite{2018MNRAS.479.2202H, 2019PASJ...71...59M, 2016SSRv..202..159T}, and gravitational waves from the merger of Pop III black holes remnants \cite{2016MNRAS.461.2722I,2020ApJ...903L..21S, 2024MNRAS.534.1634L}. Another approach, known as stellar archaeology, is to study the chemical signatures left behind by Pop III supernovae ejecta in local low-mass second-generation stars \cite{2015ARA&A..53..631F}. 
        
An alternative technique related to stellar archaeology is to analyze the metal enrichment of absorption systems in the spectra of quasi-stellar objects (QSOs). Recently, \cite{Sodini2024} presented a high-redshift sample of absorption systems selected through the detection of the 1302~\AA~ OI line. This line is a good tracer of the neutral hydrogen density, but does not suffer from the same saturation issues as Lyman-$\alpha$ at $z \gtrsim 5$ . In \cite{Sodini2024}, they find 29 $z \gtrsim 5.4$ OI absorption systems in the E-XQR-30 sample of 42 QSO spectra prepared from XSHOOTER \cite{2011A&A...536A.105V}. For all of these systems where HI column density can be measured directly, they are found to be damped Lyman-$\alpha$ absorption systems (DLAs). Based on this they assume that the entire sample is composed of very metal-poor DLAs (VMP-DLAs), and relative elemental abundances are determined from the other available metal absorption lines. Interestingly, the relative metal abundances suggest that some of the DLAs may have gas that is primarily enriched by Pop III stars. This is indicated by their large spread in relative abundances in elements such as [C/O] and [Si/O]. A spread for Pop III dominated gas is expected to be larger than for Pop II due to their smaller stellar numbers and larger variations in initial masses and supernovae energies, as predicted in the theoretical work of \cite{Vanni2024}. Note that lower-redshift DLAs ($z\approx 3$) have also been suggested as a test for Pop III star formation \cite{2022ApJ...929..158W,2023MNRAS.525..527W}.

Two previous theoretical studies \cite{Vanni2024, Kulkarni2013} were compared with the DLA data in \cite{Sodini2024}. In \cite{Kulkarni2013}, the authors present an analytic model that follows the typical growth of dark matter halos and their formation of Pop III and metal-enriched stars. This study predicts that high-redshift DLAs are promising systems to constrain the properties of Pop III stars, but does not include effects such as Lyman-Werner (LW) feedback, the baryon-dark matter velocity, merger histories of halos, and 3D metal-enrichment/reionization feedback. The other theoretical work, \cite{Vanni2024}, is mainly concerned with the chemical enrichment of DLA-analogs. They assume that the gas is formed from a single Pop III star's supernovae ejecta and mixed with Pop II gas (with the fraction of Pop II and Pop III a free parameter in the model). 
This model does not follow the detailed cosmological process of Pop III star formation or the possibility that the metal-enriched ejecta of multiple Pop III stars mix together.

In this paper, we utilize the semi-analytic model of the first stars and galaxies from \citep{2020ApJ...897...95V} to make theoretical predictions for Pop III chemical signatures in the DLAs at $z\approx 6$ and compare these predictions to observational data from \cite{Sodini2024}. Our state-of-the-art model is based on cosmological N-body simulations and takes into account the clustered 3-dimensional (3D) positions of dark matter halos. It includes a variety of physical effects necessary to compute the abundance of Pop III DLAs not considered in the previous work. This includes feedback from LW radiation, metal enrichment of the intergalactic medium (IGM), and small-scale effects of cosmic reionization (with 3D spatial fluctuations taken into account). We find that these physical processes impact the abundance and chemical enrichment from Pop III stars at the order-of-magnitude level from our previous semi-analytic work and thus are necessary to properly interpret high-redshift DLA analogs. As discussed in detail below, we use our model to predict the carbon-to-oxygen  ratios in DLAs. We find that the high values of this ratio in the observed data can potentially be attributed to Pop III metal enrichment and that the data can put strong constraints on uncertain model parameters such as the star formation efficiency and delay time between Pop III and metal-enriched star formation. 

The remainder of this paper is structured as follows. In the next section, we review our semi-analytic model of Pop III star formation and describe how we apply it to make predictions for DLAs. In Section 3, we present the results of our model. Finally, we conclude in Section 4 with a summary and discussion of our findings. Throughout this work, we assume a $\Lambda {\rm CDM}$ cosmology with parameters consistent with \cite{2014A&A...571A..16P}: $\Omega_{\rm m} = 0.32$, $\Omega_{\Lambda} = 0.68$, $\Omega_{\rm b} = 0.049$, $h=0.67$, $\sigma_8=0.83$, and $n_{\rm s} = 0.96$. These values were chosen to match the set of N-body simulations described below and utilized in this work. 

\section{Methods}

\subsection{Semi-analytical Model}
Here we provide a brief review of key aspects of our semi-analytic model (including some new modifications). For a more complete description, see \cite{2020ApJ...897...95V} (for a related model by another group, see \cite{2022ApJ...936...45H}). The foundation of our model is the 10 N-body simulation boxes from \cite{2020ApJ...897...95V}. Each box is 3 Mpc across and contains $512^3$ particles, giving a particle mass resolution of $8,000~M_\odot$. We ran these simulations with {\sc gadget2} \cite{2001NewA....6...79S} and generated dark matter halo merger trees with the publicly available codes {\sc rock star} with {\sc consistent trees} \cite{2013ApJ...762..109B, 2013ApJ...763...18B}. {\sc rock star} has been shown to accurately determine halo masses and locations with $\gtrsim 20$ particles. This corresponds to a halo mass of $M_{\rm h}\approx 2\times 10^5 ~ M_\odot$ for our simulations (which is sufficient to resolve halos hosting Pop III star formation).

Our model follows halos within the simulation boxes and applies Pop III star formation on pristine (or very low metallicity) halos when they reach the critical mass for Pop III star formation, $M_{\rm crit}$. The total Pop III mass formed in a halo is given by an efficiency, $f_{\rm III}$, yielding a stellar mass of  $M_{\rm III} = f_{\rm III} (\Omega_{\rm b}/\Omega_{\rm g}) M_{\rm h}$.
An improvement of the model in this work is an updated prescription for $M_{\rm crit}$ calibrated with the simulations of \cite{2021ApJ...917...40K}. We use their fitting formulae which include the simultaneous impact of the LW intensity, redshift, and the baryon-dark matter streaming velocity ($v_{\rm bc}$) \cite{2010PhRvD..82h3520T,2012MNRAS.424.1335F}. We note that for $z>20$ and very low values of $v_{\rm bc}$ and $J_{\rm LW}$, $M_{\rm crit}$ may go slightly lower than the resolution of our N-body simulation. However, we find that this does not significantly impact our results, as such low velocity streaming regions are rare and the DLAs are observed at $z\approx 6$. We have verified this by running a case where $M_{\rm crit}$ is set to have an artificial floor of $4\times 10^5~M_\odot$ and find that our results are essentially unchanged.

After Pop III star formation occurs in a halo, metal-enriched star formation (we do not make the distinction between Pop II/I) is assumed to follow with an efficiency $f_{\rm II}$ at a time $t_{\rm delay}$ after the Pop III burst (subsequent incoming gas also forms stars at this efficiency). The delay time is intended to model supernova feedback \cite{2014MNRAS.444.3288J}, and the star formation efficiency is set to match the UV luminosity function of high-redshift galaxies at $z\sim6$ \cite{2015ApJ...803...34B}. The model also includes ``external enrichment'' where halos enrich neighboring halos via metal bubbles produced by supernovae winds. This prescription is described in detail in \cite{2020ApJ...897...95V}. Essentially, bubbles are launched with a velocity of $v_{\rm bub} = f_{\rm bub}60~{\rm km~s^{-1}}$ and subsequently grow until they reach a comoving radius of $R_{\rm bub} = f_{\rm bub}150~h^{-1}~{\rm kpc}$ (throughout this work we assume $f_{\rm bub}=1$). The metallicity from these bubbles is tracked throughout the IGM and halos form metal-enriched stars rather than Pop III if the metallicity exceeds the critical value for Pop III star formation, $Z_{\rm crit}$ \cite{2003Natur.425..812B,2005ApJ...626..627O,2007ApJ...661L...5S,2009ApJ...691..441S}. This can trigger star formation in the low-mass halo where cooling was suppressed by LW radiation before metal enrichment. 
Larger $f_{\rm bub}$ results in more halos being externally enriched and a reduction in the amount of Pop III star formation (see Figure 6 in \cite{2020ApJ...897...95V}). We note that our choice of $f_{\rm bub}=1$ is in good agreement with the \emph{Renaissance Simulations} \cite{2015ApJ...807L..12O}.

In addition to the 3D metal enrichment of the IGM, our model also tracks 3D fluctuations in the LW flux caused by inhomogeneous star formation (as well as the smooth background LW flux generated on scales larger than the box). This is incorporated in the $M_{\rm crit}$ value of each halo. We also track the 3D ionization state of the IGM and within ionized regions halos below a characteristic mass, $M_{\rm ion}$, have their star formation suppressed due to photoheating of the IGM \cite{1998MNRAS.296...44G, 1994ApJ...427...25S,2004ApJ...601..666D,2013MNRAS.432L..51S, 2014MNRAS.444..503N}.
The various parameters of our model are listed in Table \ref{table} along with the fiducial values from \cite{2020ApJ...897...95V}.

\begin{table}
\centering
\caption{\label{table} Key physical parameters entering the semi-analytic model and their fiducial values. 
}
\begin{tabular}{c| l| c  }
Parameter & Description & Fiducial Value  \\ 
\hline
$f_{\rm III}$ & Pop III star formation efficiency  &  0.001 \\
$f_{\rm II}$ & Metal-enriched star formation efficiency &  0.05 \\
$t_{\rm delay}$ & Delay in subsequent star formation due to Pop III SNe &  $10^7$ yr  \\
$Z_{\rm crit} $ & Critical metallicity for externally metal-enriched halos  &  $3 \times 10^{-4}~Z_\odot$ \\
$M_{\rm min, met} $ & Critical mass for externally metal-enriched halos  &  $2\times 10^5~M_\odot$  \\
$M_{\rm ion} $ & Ionization feedback mass &  $1.5\times 10^8  \left ( \frac{1+z}{11} \right )^{-3/2}~M_\odot$  \\
$f_{ \rm esc, II}$ & Ionizing escape fraction in metal-enriched halos  &  0.1  \\
$f_{ \rm esc, III}$ & Ionizing escape fraction in Pop III halos  &  0.5   \\
$v_{\rm bc}$ & Streaming velocity at recombination &  $0-60~{\rm km~s^{-1}} ~ (0-2\sigma)$   \\
$f_{\rm bub}$ & Metal bubble size scaling factor &  1  \\[1ex]
$M_{\rm min}$ &  Minimum halo mass for Pop III star formation   &   fits from \cite{2021ApJ...917...40K}  \\
\end{tabular}
\end{table}

\subsection{Abundance of Pop III DLAs}
We utilize our model to make predictions for $z {\approx} 6$ DLAs and compare them with the observations of \cite{Sodini2024}. In order to compare with the data, we make several simplifying assumptions. First, we assume that half of the metals created stays in the halos, and the other half is ejected into the IGM. 
While this is a very rough approximation, we emphasize that the elemental abundance ratios from a mixture of Pop III and metal-enriched stars do not directly depend on this assumed fraction. It is true, however, that as this fraction escaping into the IGM goes down, external metal enrichment is reduced (metal pollution from one halo to a neighboring one). In our model, this effect is degenerate with the critical metallicity for Pop III star formation, $Z_{\rm crit}$ (which we vary in our analysis below).
We also assume that DLAs probe gas within dark matter halos and that within halos the metals are well mixed.

\begin{figure}
\centering 
\includegraphics[width=12 cm]{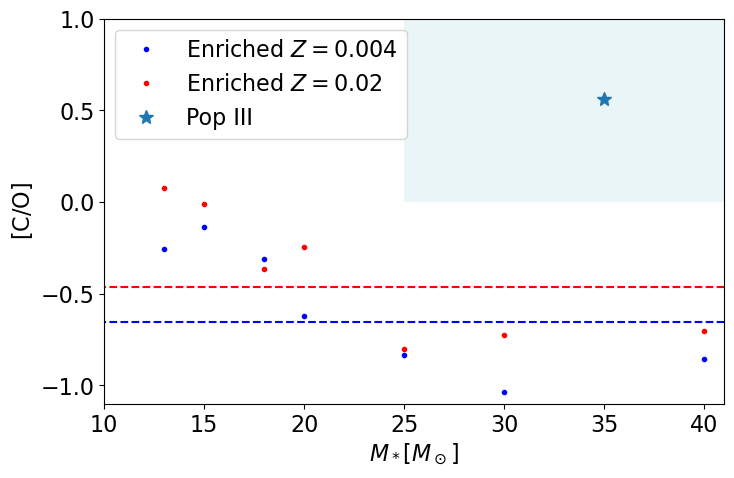}
\caption{\label{fig:stars_CO} The carbon-to-oxygen elemental abundance ratio of supernovae ejecta as a function of stellar mass for our assumed metal yields. Metal-enriched yields are assumed to follow \cite{2006ApJ...653.1145K} with a fixed metallicity of the stars (either $Z=0.004$ or $Z=0.02$). Pop III yields are taken from \cite{2023MNRAS.526.4467J} with $v_{\rm rot}/v_{\rm crit} = 0.6$. There is only one Pop III point due to the simplifying assumption that all Pop III stars have the same metal yields. 
 The shaded rectangle around the Pop III point is included to emphasize the general uncertainty for Pop III stars (but we note that the [C/O] values do not change by more than roughly ten percent in the specific case of rapidly rotating Pop III stars from Tables 2 and 3 in \cite{2023MNRAS.526.4467J}, which vary stellar mass and rotation speed).
The dashed lines indicate the [C/O] produced by a massive cluster of metal-enriched stars, sampling over our assumed Kroupa IMF. }
\end{figure}

We assume that the majority of Pop III stars are rapidly rotating, which \cite{2023MNRAS.526.4467J} show can lead to the metal yields found in carbon-enhanced metal-poor (CEMP) stars \cite{2005ARA&A..43..531B, 2007ApJ...655..492A}. 
There are several different classes of CEMP stars, with CEMP-no having very low abundance of heavy elements and likely originating from metals created by Pop III stars \cite{2013A&A...552A.107S, 2013ApJ...762...28N}. Proposed origins for CEMP-no stars include both faint supernovae \cite{2003Natur.422..871U, 2010ApJ...724..341H} and rapidly rotating Pop III stars (as we focus on here).
For simplicity, we assume that all Pop III stars in our semi-analytic model have a stellar mass of $35~M_\odot$. Yields are taken from \cite{2023MNRAS.526.4467J} with $v_{\rm rot}/v_{\rm crit} = 0.6$ (see their Table 3).
These Pop III stars likely do not go supernovae, and instead their metals are dispersed via stellar winds roughly two orders of magnitude less energetic than core collapse supernovae \cite{2023MNRAS.526.4467J}. 
However, we note that our model assumes a delay time between Pop III and Pop II star formation caused by energetic supernovae. 
Thus, we assume that while the majority of the Pop III metal enrichment is caused by stellar winds from rapidly rotating stars, our halos are subjected to a small number of
supernovae from non-rotating Pop III stars (which for simplicity, we do not include metals from). We note that for our fiducial Pop III star formation efficiency, a typical atomic cooling halo at $z\approx 6$ will make $\sim 10^4~M_\odot$ of Pop III stars in a starburst, so it would be possible in principle to have a small fraction of $\lesssim 100 ~M_\odot$ Pop III stars ending in supernovae which contribute energetically, but have a small impact on metal enrichment. We assume these supernovae contribute to half the metals being ejected from the halos into the IGM. We have made this simplified assumption because the rapidly rotating Pop III stars \cite{2023MNRAS.526.4467J} have yields with [C/O]$\sim0.6$, which is approximately as high as the DLA analogs from \cite{Sodini2024} extend to. We note that for non-rotating Pop III core collapse supernovae modeled in \cite{2010ApJ...724..341H}, the values of [C/O] vary over four orders of magnitude depending on uncertain parameters associated with the supernovae explosion (e.g., energy). Many of these produce very low yields of C and O, or [C/O]$<0$. The pair instability models from \cite{2002ApJ...567..532H} have [C/O]<0 and thus could not explain the high-[C/O] DLAs. We emphasize that our main results presented below depend strongly on the assumed abundances. We defer a comprehensive study of the range of possible yields for future work.
For metal-enriched stars, we assume a Kroupa IMF \cite{2001MNRAS.322..231K} over a stellar mass range of $0.01-100~M_\odot$ (see Eq.~5 in \cite{2022ApJ...936...45H}) and metal yields given by \cite{2006ApJ...653.1145K} assuming a fixed metallicity of the stars as either $Z=0.004$ or $Z=0.02$.
In Figure \ref{fig:stars_CO}, we show the [C/O] produced in ejected metals as a function of mass for both Pop III and metal-enriched stars. 

When computing the abundance of DLAs, we take the baryon-dark matter streaming velocity ($v_{\rm bc}$) into account \cite{2010PhRvD..82h3520T,2012MNRAS.424.1335F}. 
The values of $v_{\rm bc}$ throughout the universe follow a Maxwell-Boltzmann distribution, $f_{\rm bc}(v_{\rm bc}) \propto v_{\rm bc}^2 \exp\left [ {-3v_{\rm bc}^2/(2 \sigma_{\rm bc}^2)} \right ] $, where $\sigma_{\rm bc}$ is the root mean square value of the velocity.
Note that the value of $v_{\rm bc}$ does not vary significantly across 3 Mpc scales, so each of our ten simulation boxes is given a constant value of $v_{\rm bc}$. One box has a value of $0.25 \sigma_{\rm bc}$, two have values of $0.5 \sigma_{\rm bc}$, four have values of $1 \sigma_{\rm bc}$, two have values of $1.5 \sigma_{\rm bc}$, and one has a value of $2 \sigma_{\rm bc}$. We compute the global abundance of DLAs by assuming each of our halos represents a set of similar halos throughout a much larger volume. Thus, for each \emph{halo} in our model, we assign an effective $v_{\rm bc}$-weighted number density of
\begin{equation}
n_{j,k} =  \frac{ f_{\rm bc}(v_{{\rm bc,} j}) }{V_{\rm box}\sum_{i} f_{\rm bc}(v_{{\rm bc,} i}) },
\end{equation}
where the sum is over the ten simulation boxes, $v_{{\rm bc,} i}$ is the streaming velocity in the $i^{\rm th}$ box, $j$ is the box which contains the halo whose number density is being computed, $k$ is an index numbering the specific halo, and $V_{\rm box}$ is the comoving volume of each simulation box.

We note that the LW radiation seen by each halo is due both to a small-scale spatially-fluctuating component from sources in the box as well as a large-scale component from distances much larger than the box (with a LW horizon of $\approx 100 ~ {\rm Mpc}$). In order to treat this background most accurately, for each model parameterization we test below, we first perform a preliminary run to estimate the large-scale global component of $J_{\rm LW}(z)$ (which is then applied to all of the boxes in the final runs). For the preliminary run, we assume $v_{\rm bc} = 1\sigma_{\rm bc}$, which as shown in \cite{2024ApJ...962...62F} is expected to give very similar values as one would get averaging over $f_{\rm bc}(v_{\rm bc})$. For redshifts of $z>30$ in the preliminary run, we assume a global $J_{\rm LW}(z)$ equal to the $f_{\rm III } = 0.001$ case presented in Appendix A of \cite{2024ApJ...962...62F} (though we find our results are insensitive to this choice). For lower redshifts in the preliminary run ($z<30$), we compute the global $J_{\rm LW}(z)$ self-consistently using the star formation history of each box as described in \cite{2020ApJ...897...95V}.  

Below we predict the expected number of DLAs found in various [C/O] bins for each QSO spectrum. This is well approximated by 
\begin{equation}
\label{eq:N_DLA}
N_{\rm DLA} \approx \sum_j \sum_k \Delta L_j ~ n_{k} A_{\rm eff}(M_{{\rm h}},z_j),
\end{equation}
where $N_{\rm DLA}$ is the mean number of DLAs observed in some [C/O] bin, $n_{k}$ is the effective number density of halo $k$, $\Delta L_j$ is the comoving distance between redshifts $z_j$ and $z_{j-1}$, and  $A_{\rm eff}(M_{{\rm h}},z_j)$ is the effective area of a halo that surpasses the DLA column density threshold. The sums in the equation are over our simulation snapshot redshifts (indexed by $j$) and over all of the halos in all 10 boxes that fall into the particular [C/O] bin under consideration (with halos indexed by $k$). Our simulation snapshots extended down to a redshift of $z=5.9$ and are spaced in time by two percent of the age of the universe at each snapshot redshift.

We note that while equation \ref{eq:N_DLA} appears simple, it implicitly contains a substantial amount of physics, as followed by our semi-analytic model. For each halo, $n_k$ only contributes to the [C/O] bin corresponding to that halo's [C/O]. Each halo's elemental abundance ratios are determined by the amount of Pop III and metal enriched stars formed in the halo which depends on a variety of important feedback processes (e.g., LW feedback, external metal enrichment, and baryon-dark matter streaming velocity) and the halo's unique merger history and environment (e.g., local LW background and reionization history).

We also note that this abundance estimate does not take into account the spatial clustering of DLA-hosting halos (though clustering is taken into account when simulating the galaxy properties). In future work, we will create mock data by tracing lines of sight through our simulation box to capture clustering effects. While this would not alter our mean abundance of DLAs, it could impact the variance among different quasar spectra.

We approximate the effective area following \cite{Kulkarni2013} who utilize a calibration from the $z=3$ simulations of \cite{2008MNRAS.390.1349P}. Thus, we assume 
\begin{equation}
\label{eqn:A_eff}
A_{\rm eff}(M_{\rm h}) = \epsilon_0 \left ( \frac{M_{\rm h}}{M_0} \right )^2 \left ( 1 +  \frac{M_{\rm h}}{M_0} \right )^{\alpha-2},
\end{equation}
where $\epsilon_0= 40(1+z)^2~{\rm kpc}^2$, $\alpha = 0.2$, and $M_0=10^{9.5}~M_\odot$ at $z=3$. $M_0$ is assumed to scale with redshift such that the halo circular velocity is constant (i.e., $M_0 \propto (1+z)^{-1.5}$). 
The physics driving this equation is that more massive halos can retain more photoionized gas and promote more efficient radiative cooling (leading to more neutral hydrogen in the halo). We note that $M_0$ corresponds roughly to the Jeans mass for gas photoheated by ionization. Below this mass, the DLA effective area drops off very rapidly with mass ($A_{\rm eff}\propto M_{\rm h}^2$) since halos cannot retain ionized hydrogen. Above this, there is a much weaker dependence ($A_{\rm eff}\propto M_{\rm h}^{0.2}$).

As described below, we find that increasing $A_{\rm eff}$ by a factor of $\sim 4$ is required to match the observed abundance of DLA analogs in \cite{Sodini2024}. It is plausible that the lower ultraviolet background at higher redshift ($z{\approx}6$) results in increased neutral gas and thus a higher $A_{\rm eff}$. We note that this is a rough approximation which has a strong impact on our results and motivates additional hydrodynamical cosmological simulations in future work.

\section{Results}
\label{sec:results}

\begin{figure}
\centering 
\includegraphics[width=12 cm]{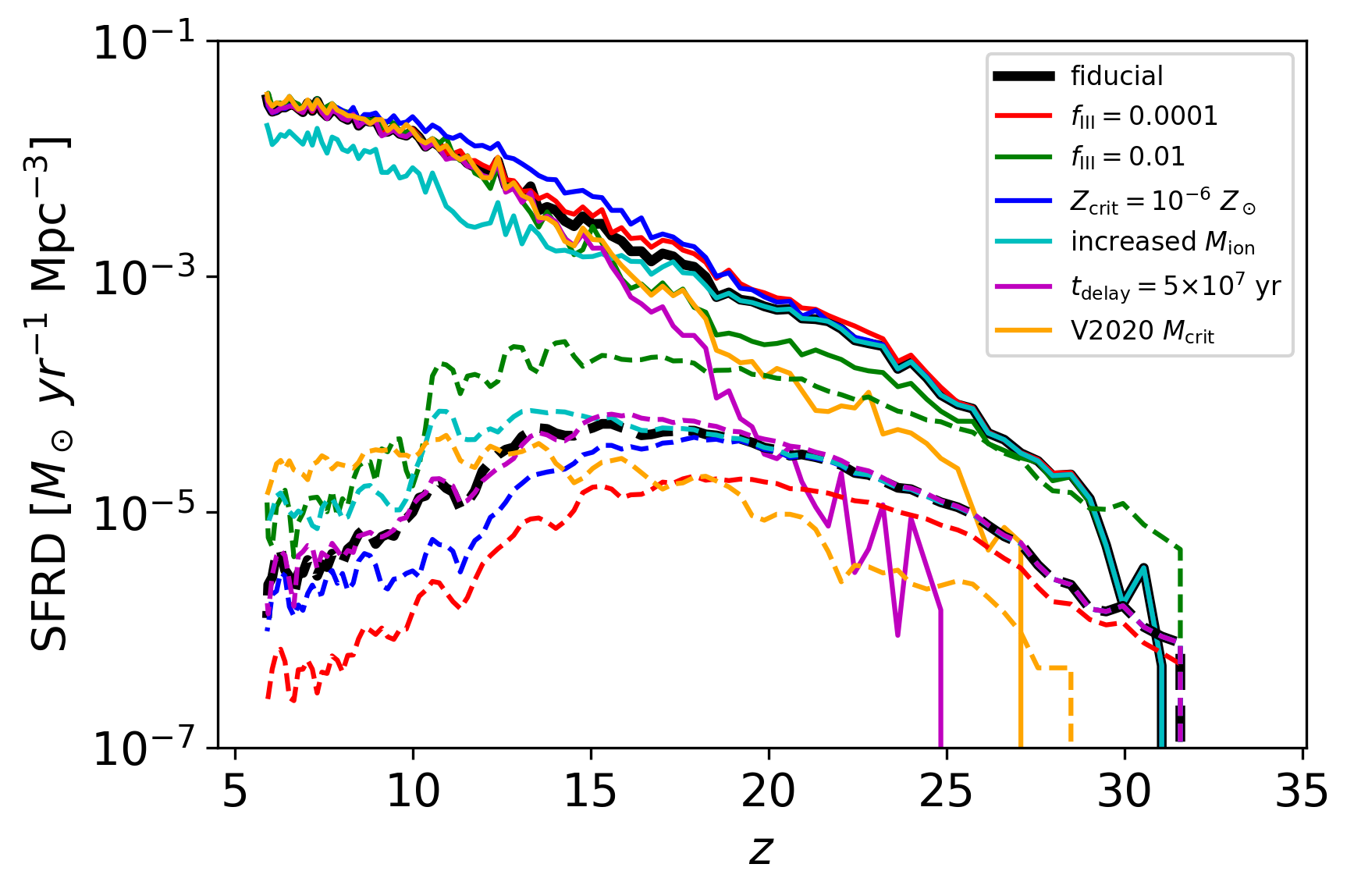}
\caption{\label{fig:SFRDs} The metal-enriched (solid curves) and Pop III (dashed curves) SFRDs computed from our 10 N-body simulations (using an average weighted by their assigned $f(v_{\rm bc})$ values). The fiducial model parameters are listed in Table \ref{table}. For the other curves, only one parameter is varied from the fiducial values as indicated in the legend. ``V2020 $M_{\rm crit}$'' uses the formula from \cite{2020ApJ...897...95V} (based on simulations of \cite{2001ApJ...548..509M}), and the rest of the simulations use the fitting form from the more recent simulation of \cite{2021ApJ...917...40K}. The curve labeled ``increased $M_{\rm ion}$'' has been assigned $5\times 10^8  \left ( \frac{1+z}{11} \right )^{-3/2}~M_\odot$.
}
\end{figure}

The main goal of this paper is to determine whether our semi-analytic model predicts Pop III chemical signatures in $z\approx6$ DLAs similar to those presented in \cite{Sodini2024} and to see if these data have the power to constrain uncertain model parameters such as the Pop III star formation efficiency. To address this, we explore predictions for a range of model parameterizations. We begin with the fiducial model summarized in Table \ref{table} and vary individual parameters one at a time, but leave a comprehensive survey of the parameter space to future work. The resulting Pop III and metal-enriched star formation rate densities (SFRDs) are shown in Figure \ref{fig:SFRDs} (note that additional parameter variations were  explored in \cite{2020ApJ...897...95V}). We see that there is an order-of-magnitude variation in the Pop III SFRD, which is also reflected in the Pop III chemical signatures in DLAs explored below. 

\begin{figure}
\centering 
\includegraphics[width=12 cm]{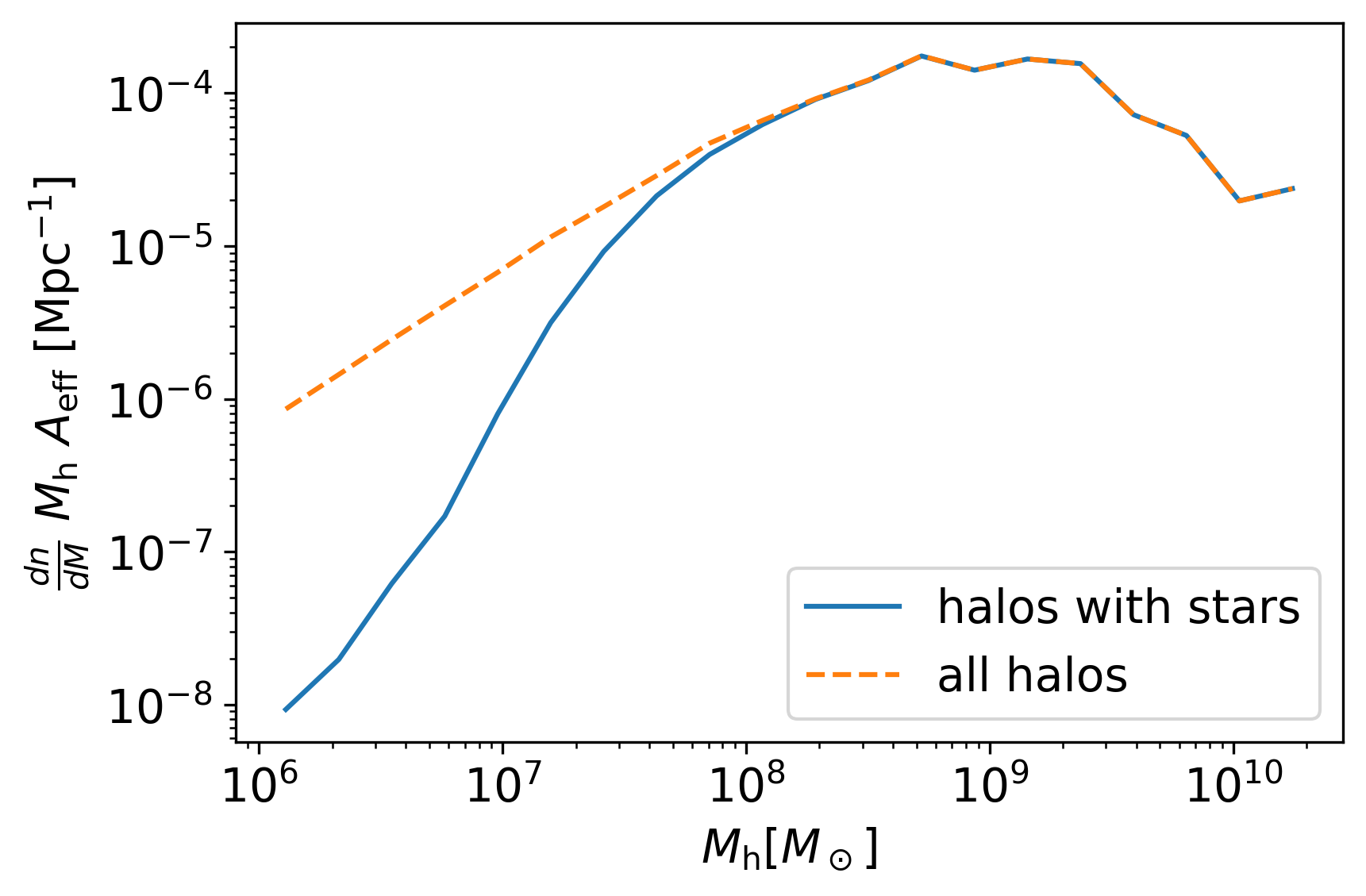}
\caption{\label{fig:nA} The covering factor per comoving Mpc along the line of sight for halos as a function of their mass at $z=5.9$ (computed from Eq.~\ref{eqn:A_eff}). This gives the relative contribution of halos with different masses to the observed DLAs. We show the halo densities with and without stars. The former are computed with our fiducial model parameterization.}
\end{figure}

Before exploring Pop III chemical signatures, we examine the relative contribution of different halos in our model to the DLA abundance. In Figure \ref{fig:nA}, we plot the product of the halo number density and $A_{\rm eff}$ as a function of halo mass at $z\approx6$ (i.e., their covering factor per comoving Mpc along the line of sight). This reflects the relative contribution of different halo masses to the expected number of DLAs. We find that most DLAs in our model are from halos with $M_{\rm h} = 10^8~M_\odot - 10^{9.5}~M_\odot$ (with $90\%$ originating from $M_{\rm h} > 10^8~M_\odot$). Lower-mass halos do not contribute strongly due to their low $A_{\rm eff}$ and higher-mass halos due to their low number density. We note that in our fiducial model there are ${\sim}10,000$ star forming halos at $z \sim 6$ across all 10 N-body simulation boxes (with ${\sim}2000$ above halo mass of $10^8~M_\odot$).

\begin{figure}
\centering 
\includegraphics[width=7 cm]{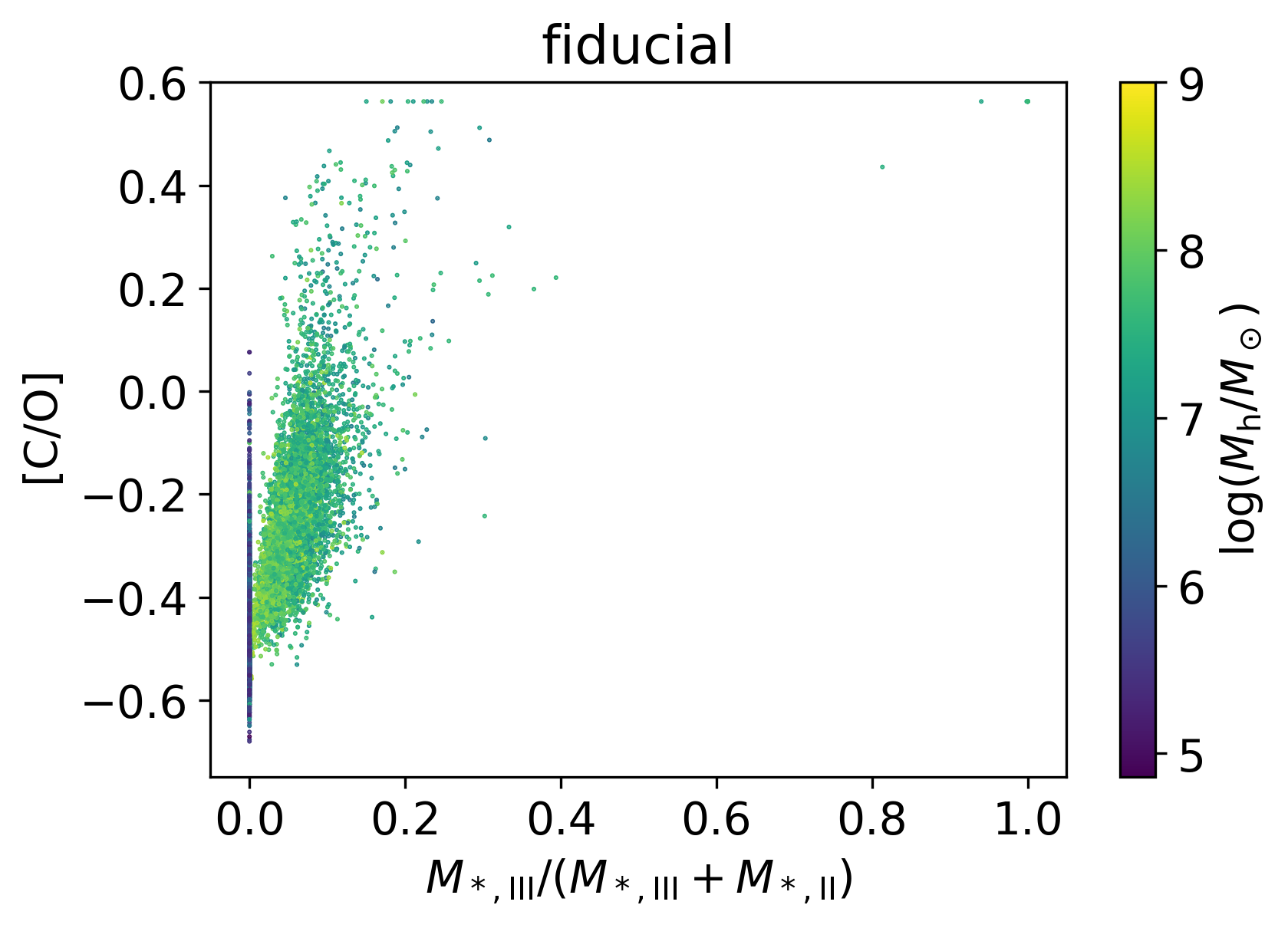}
\includegraphics[width=7 cm]{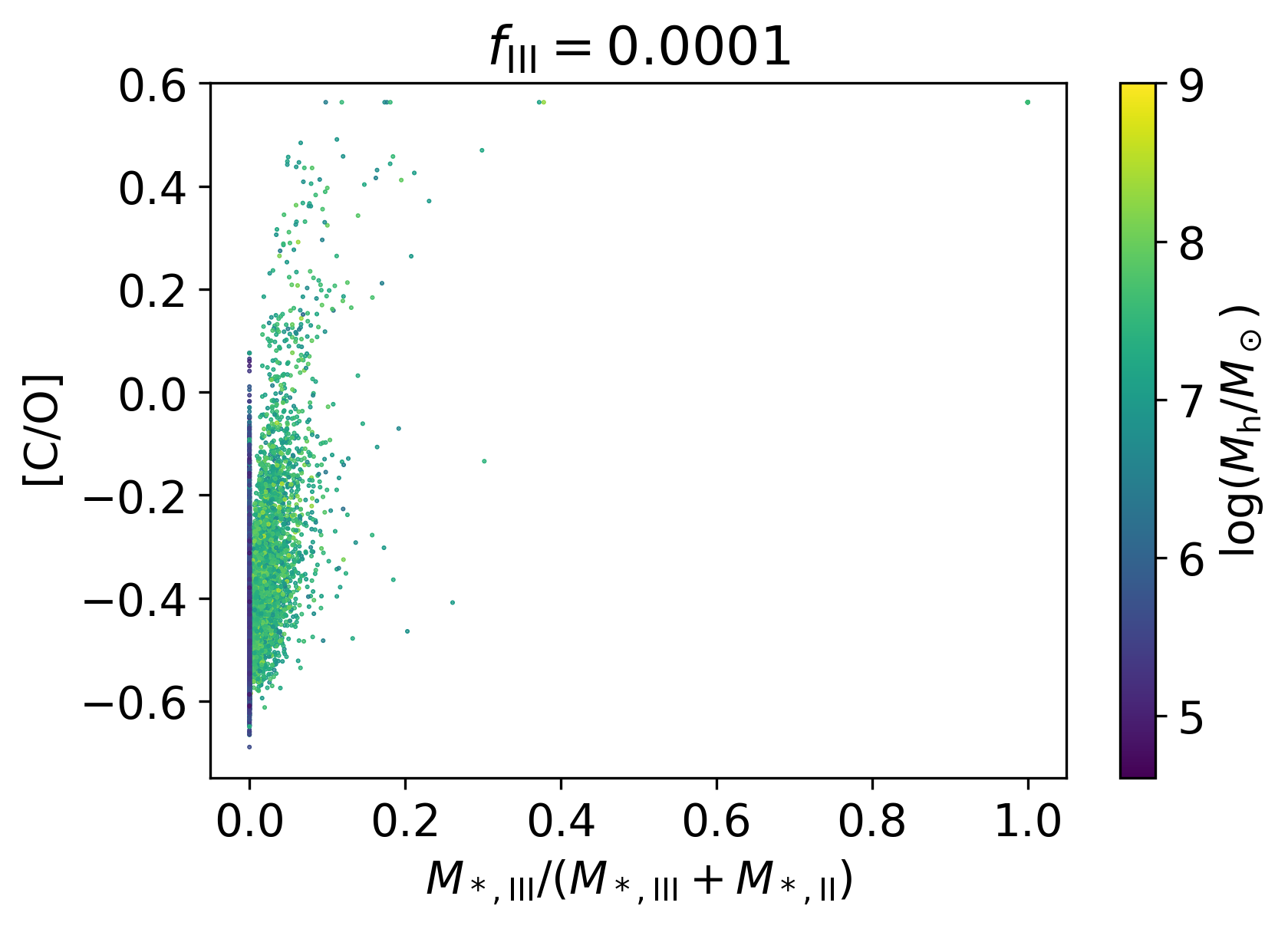}
\includegraphics[width=7 cm]{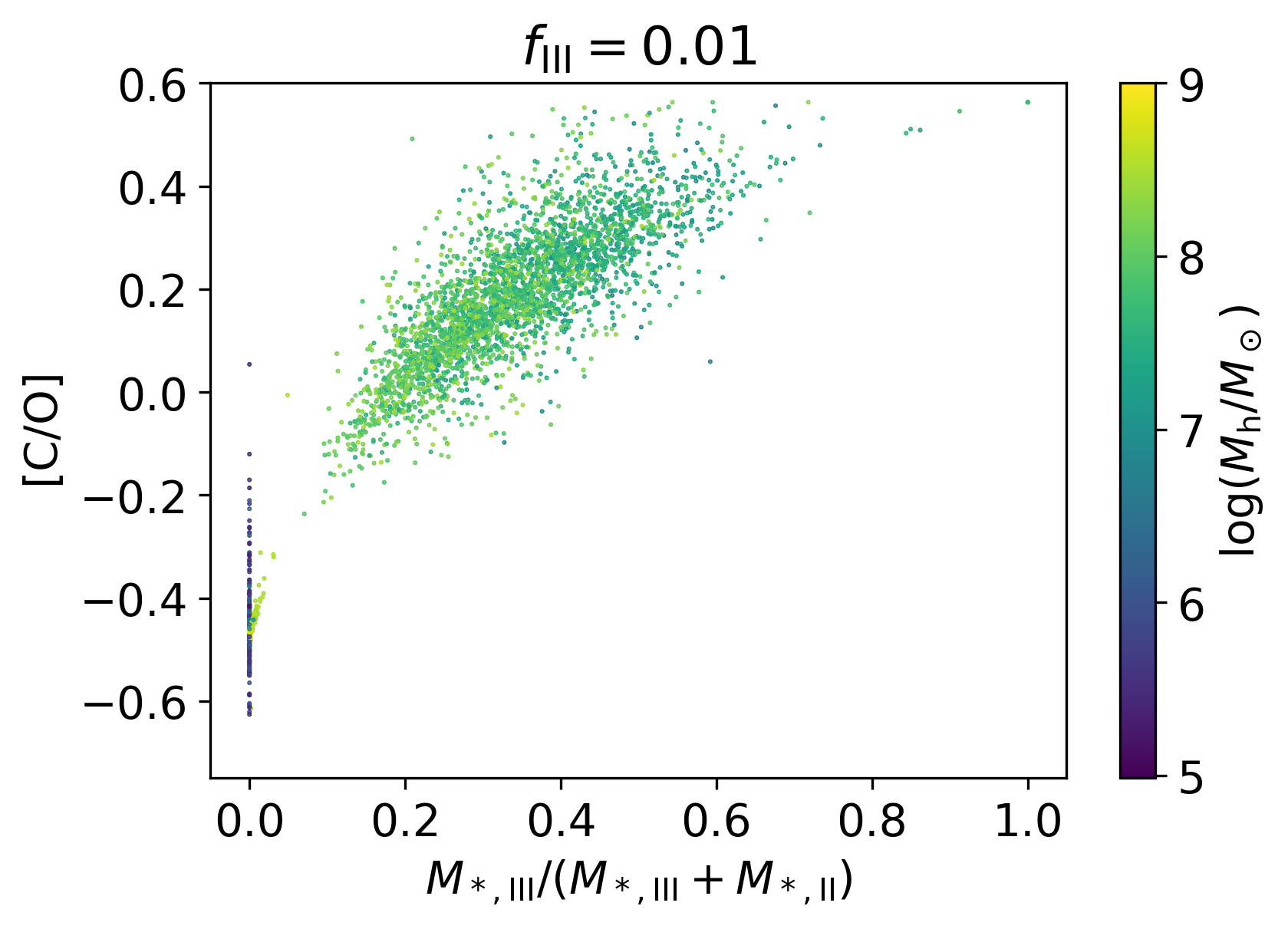}
\includegraphics[width=7 cm]{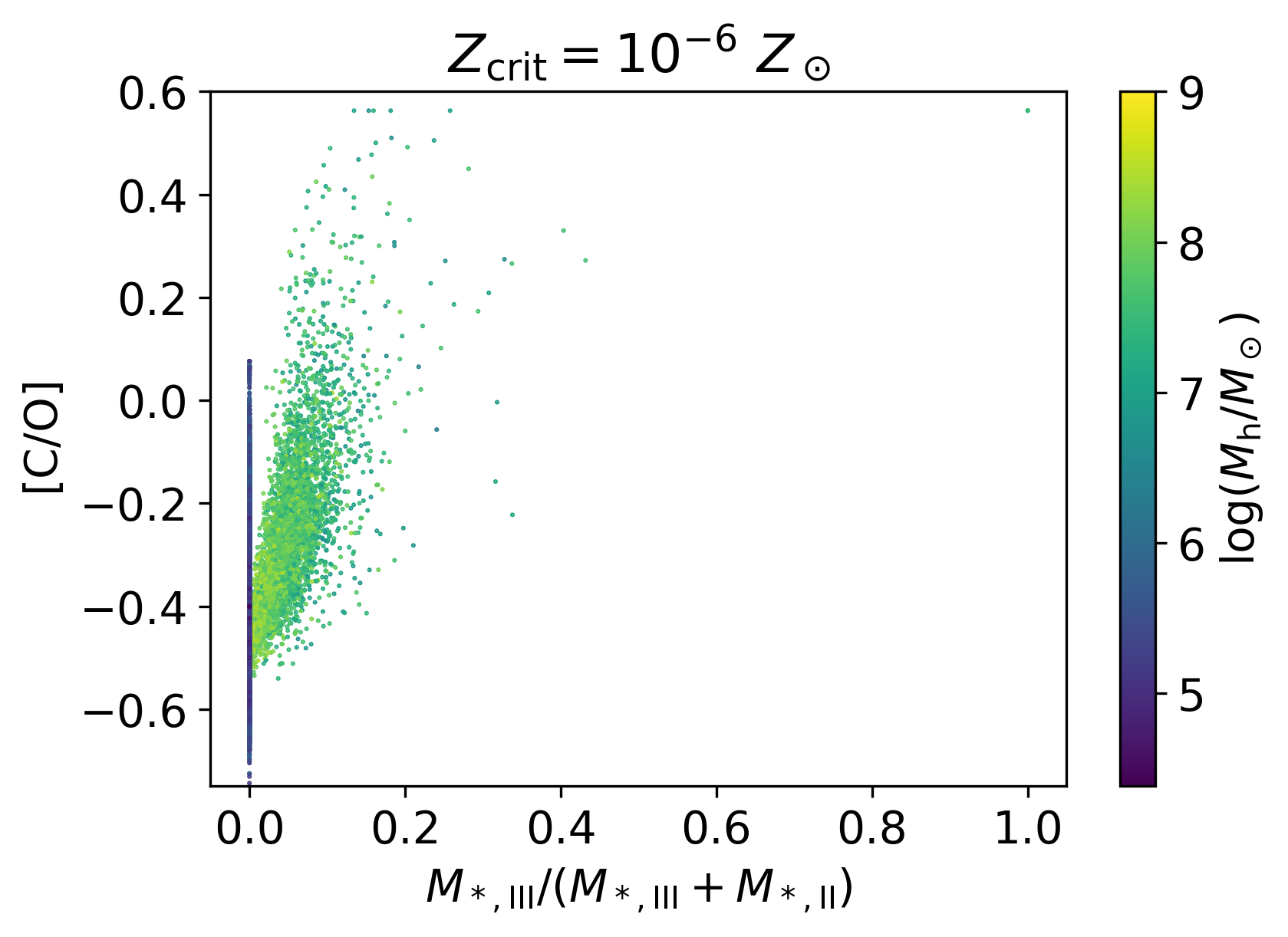}
\includegraphics[width=7 cm]{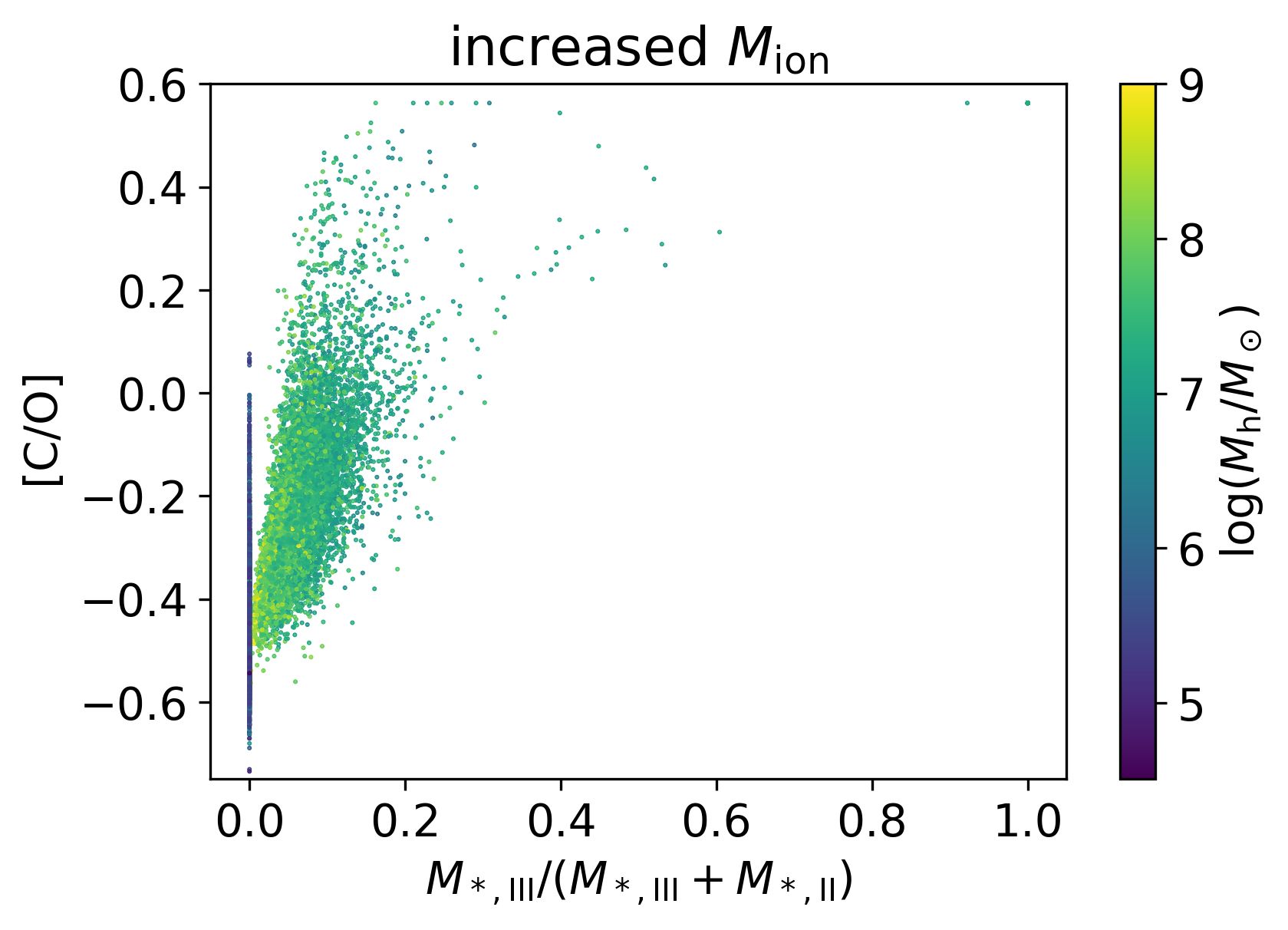}
\includegraphics[width=7 cm]{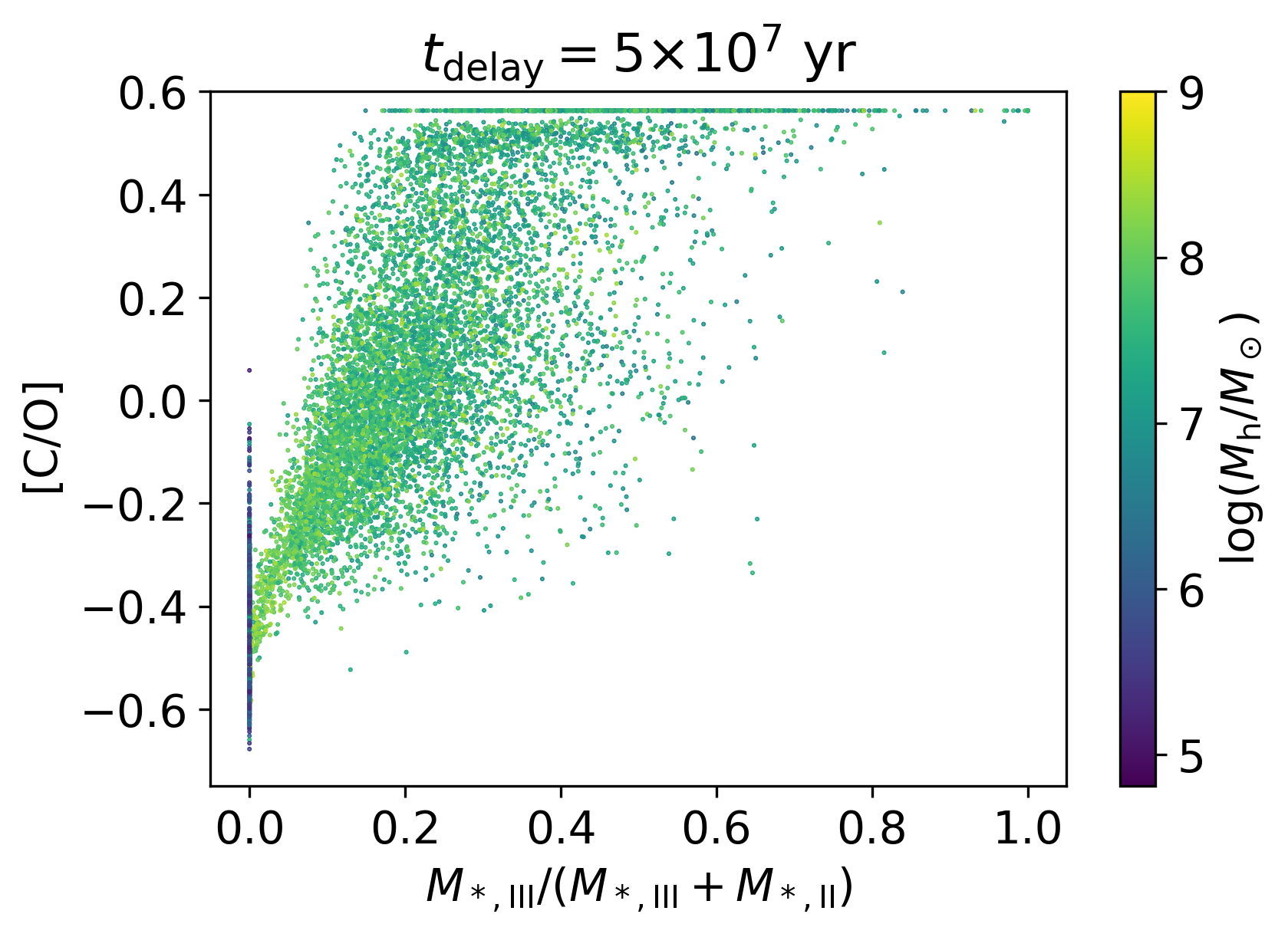}
\includegraphics[width=7 cm]{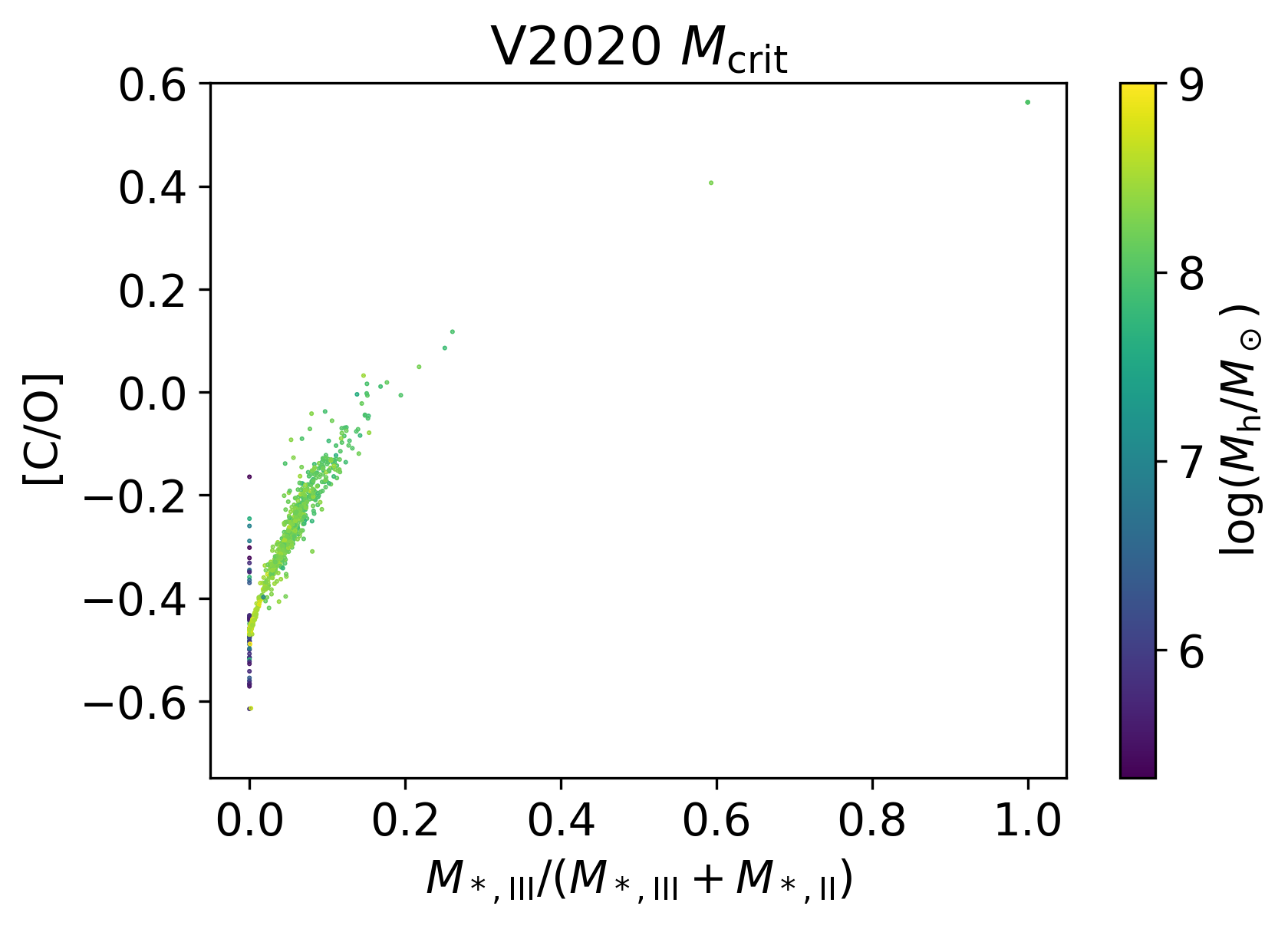}
\caption{\label{fig:scatter} Carbon-to-oxygen abundance ratios in the ISM for all of the halos in our model with stars at $z=5.9$ as a function of the ratio of cumulative Pop III mass to total cumulative stellar mass ($M_{\rm *, III}/[M_{\rm *, III}+M_{\rm *, II}]$). These masses include all stars formed within the halo or its progenitors. Colors indicate the virial mass of the halos. Results are shown for all of our single-parameter variations around the fiducial model. The ``V2020 $M_{\rm crit}$'' and ``increased $M_{\rm ion}$'' models are the same as described in the caption of Figure \ref{fig:SFRDs}. Note that all halos with star formation are included, but that masses above $10^9~M_\odot$ are suppressed in the color bar as they are all tightly clustered near zero Pop III fraction and [C/O]$\sim -0.45$.   }
\end{figure}

\begin{figure}
\centering 
\includegraphics[width=7 cm]{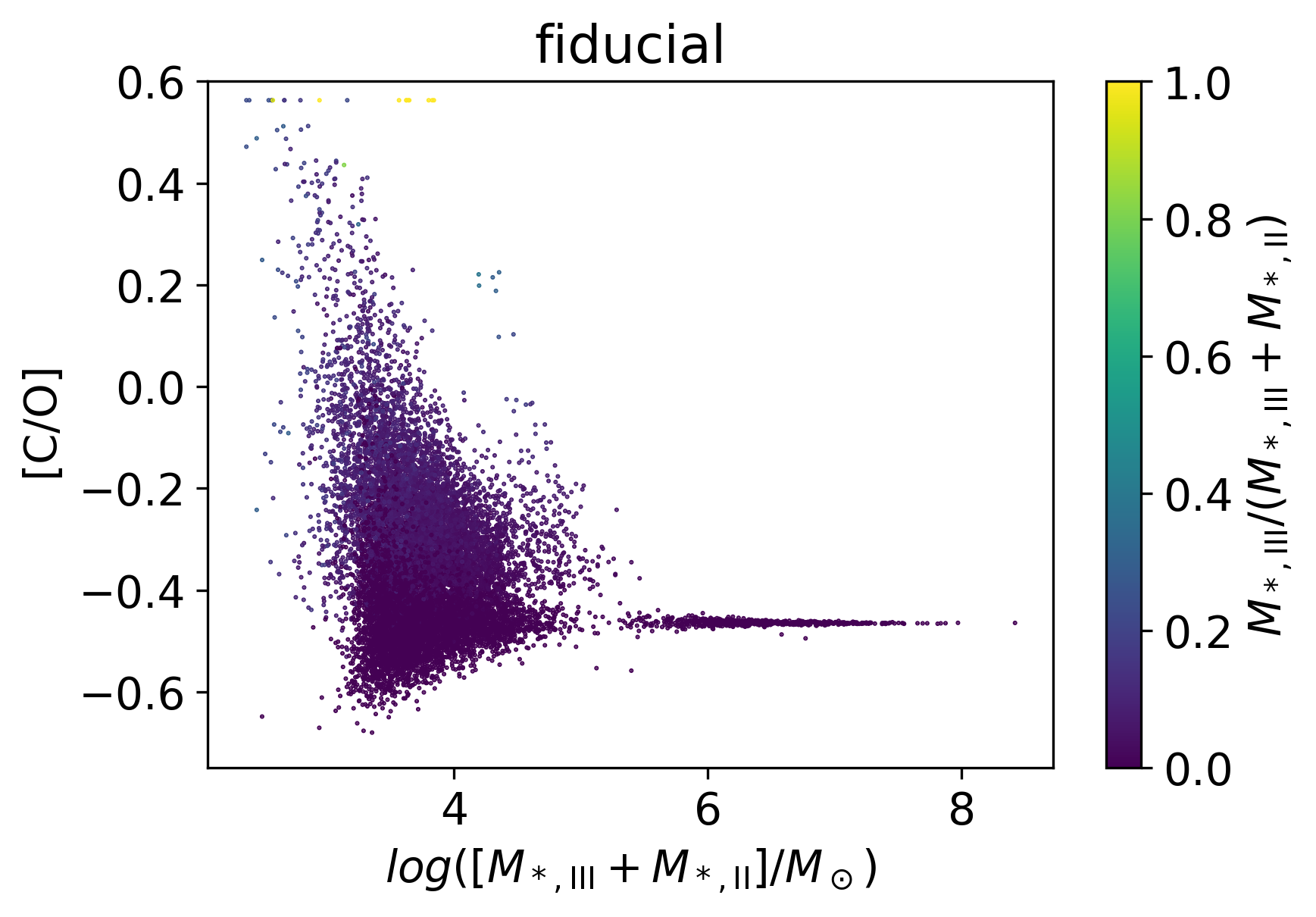}
\includegraphics[width=7 cm]{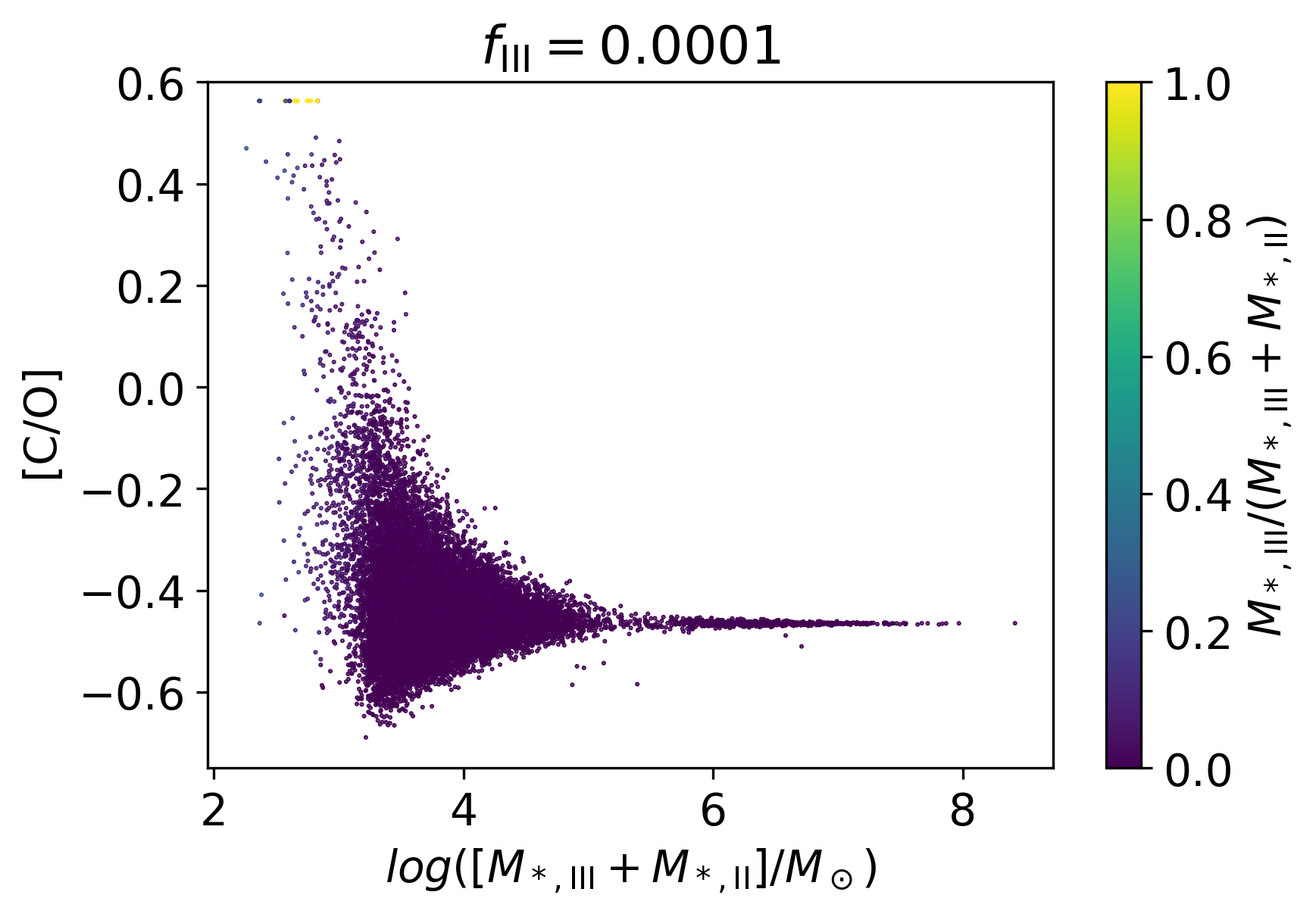}
\includegraphics[width=7 cm]{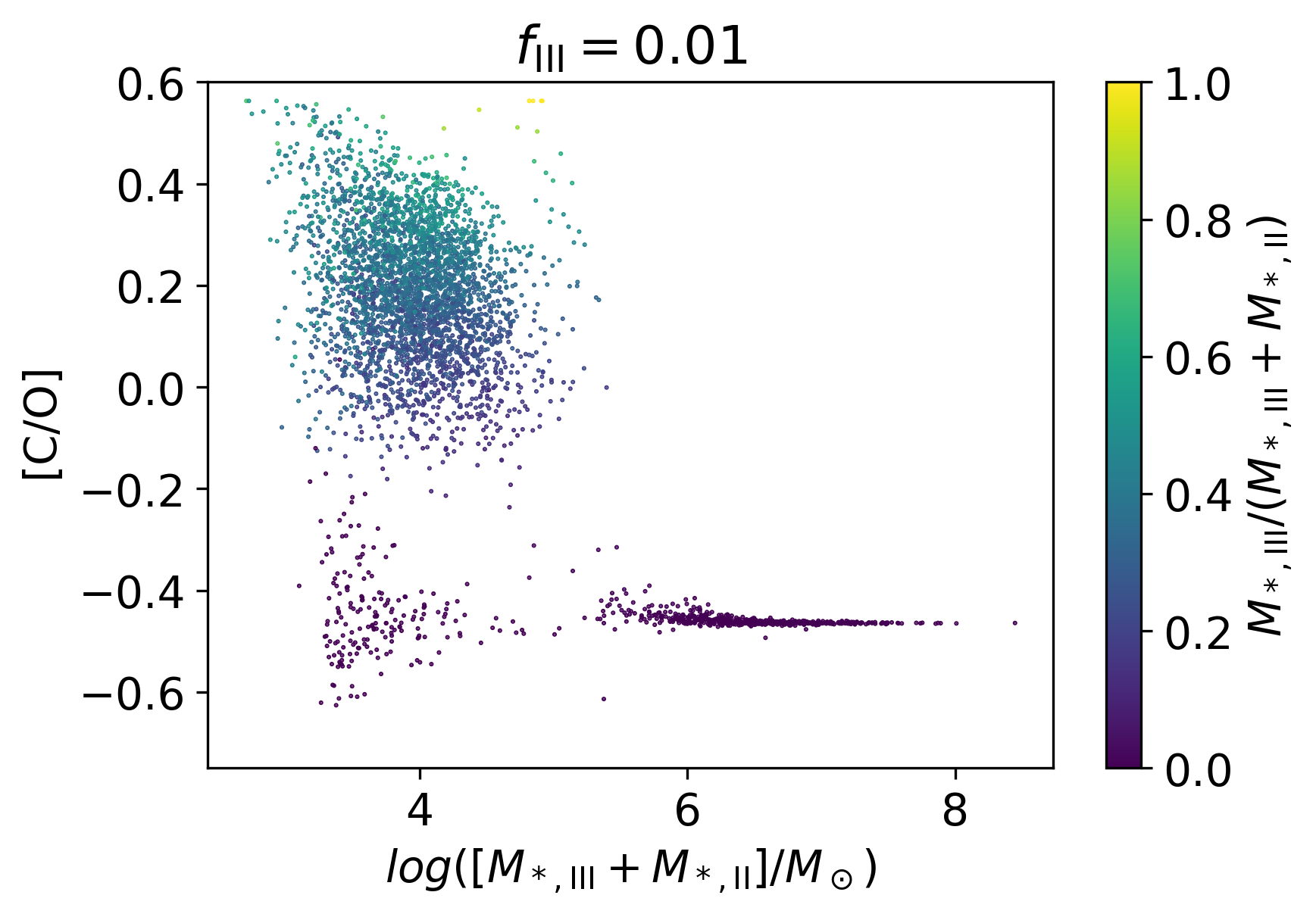}
\includegraphics[width=7 cm]{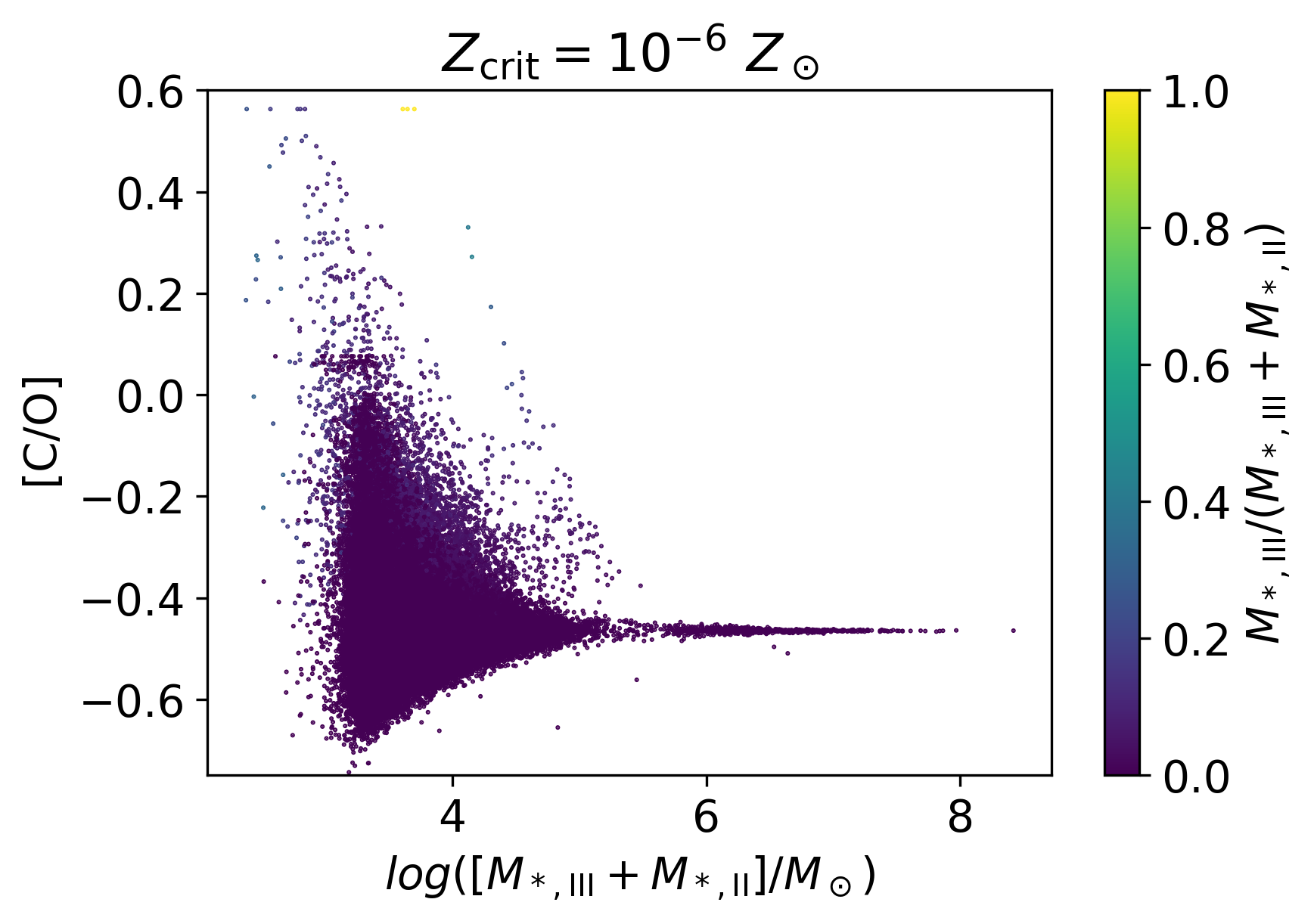}
\includegraphics[width=7 cm]{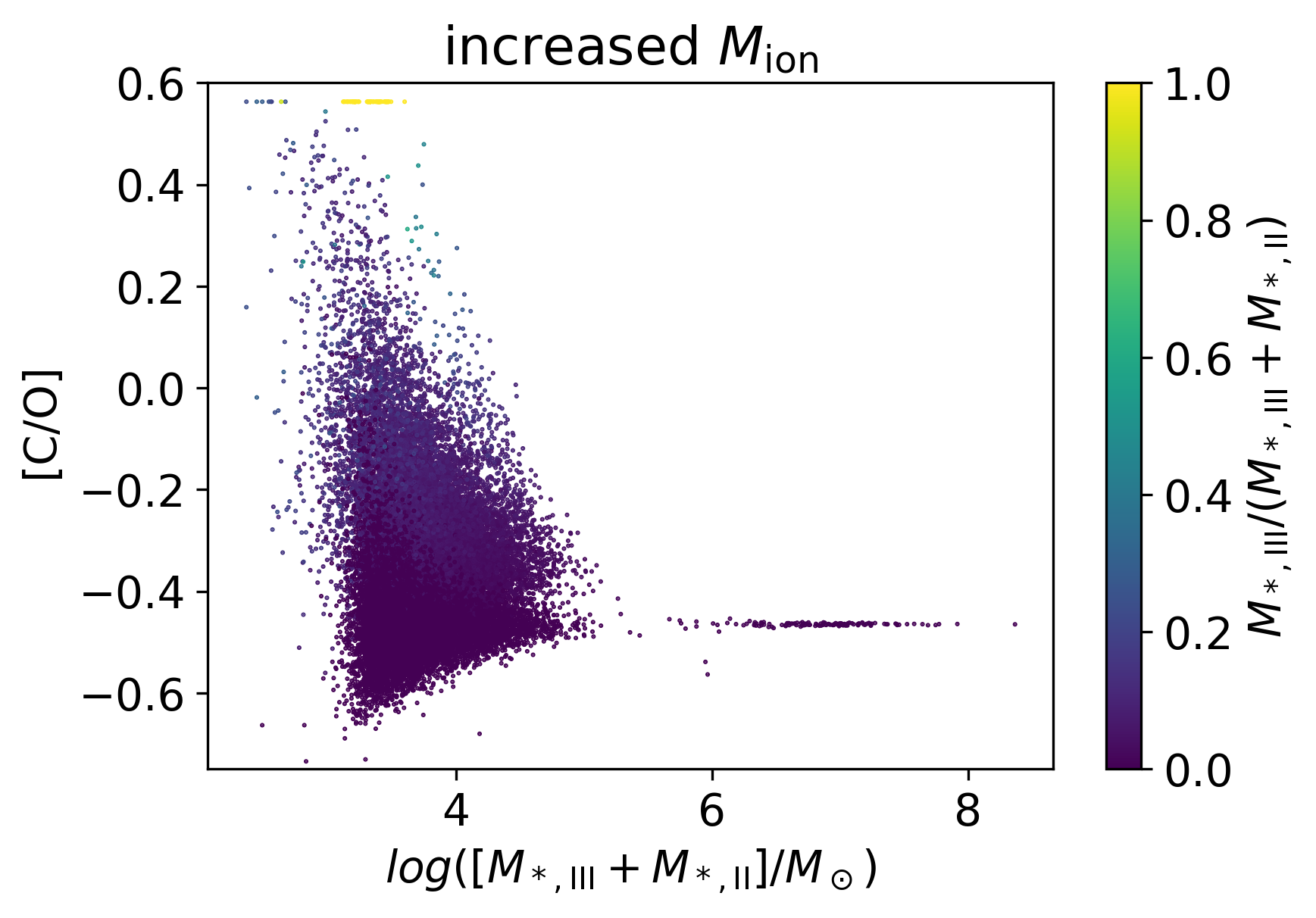}
\includegraphics[width=7 cm]{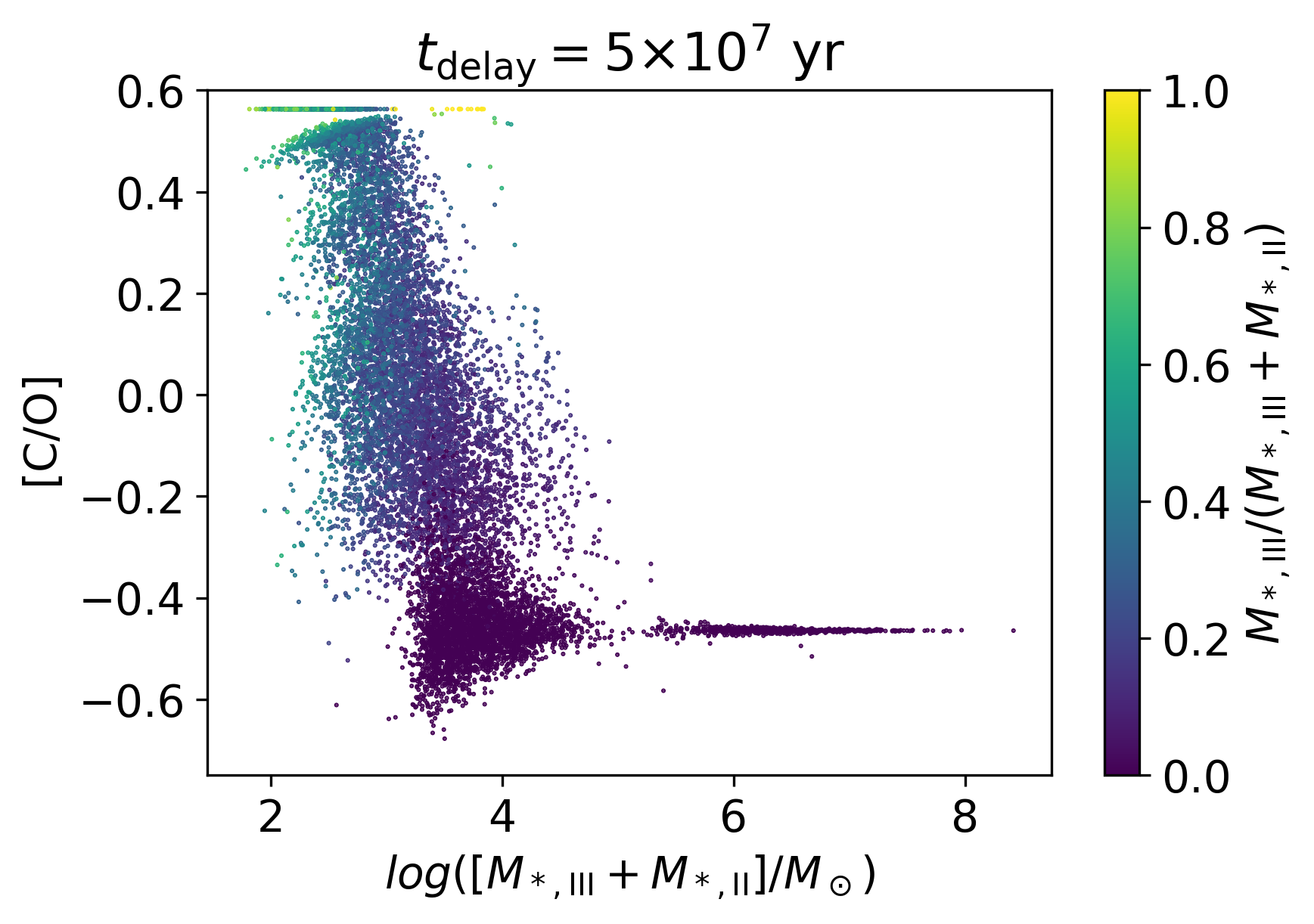}
\includegraphics[width=7 cm]{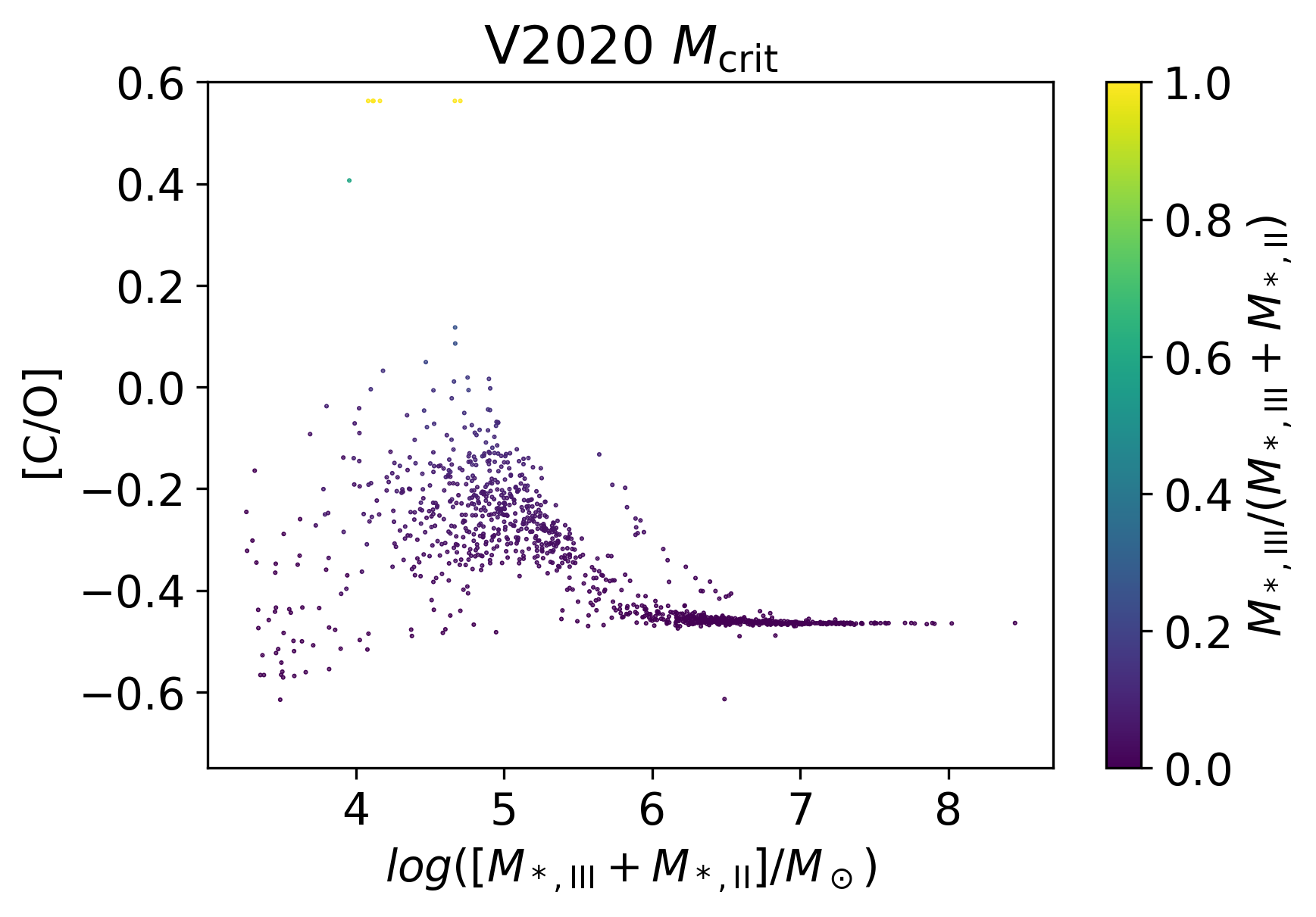}
\caption{\label{fig:scatter2} Carbon-to-oxygen abundance ratios in the ISM for all of the halos with stars in our model at $z=5.9$ as a function of total cumulative stellar mass ($M_{\rm *, III}+M_{\rm *, II}$). Colors indicate cumulative fraction of Pop III stars ($M_{\rm *, III}/[M_{\rm *, III}+M_{\rm *, II}]$). Results are shown for all of our single-parameter variations around the fiducial model. The ``V2020 $M_{\rm crit}$'' and ``increased $M_{\rm ion}$'' models are the same as described in the caption of Figure \ref{fig:SFRDs}. }
\end{figure}

We predict the [C/O] ratios in $z\approx 6$ DLAs and compare to observations later in this section. However, in order to interpret these results, we first examine how [C/O] varies  within halos in the semi-analytic model. In Figures \ref{fig:scatter} and \ref{fig:scatter2}, we show [C/O] in the entire population of star-forming halos (from all 10 N-body simulation boxes and all model parameterizations), as a function of fraction of cumulative stellar mass in Pop III stars, total cumulative stellar mass, and halo virial mass (cumulative stellar mass includes all stars ever formed in a halo and its progenitors). Note that in these figures, we have assumed $Z=0.02$ for metal-enriched stars when computing metal yields, as this agrees better with the observational data \cite{Sodini2024}, but find qualitatively similar results for $Z=0.004$. For the comparisons to observations below, we show results for both $Z=0.02$ and $Z=0.004$ which bracket reasonable values for our high-redshift galaxies.

There a number of interesting features in these plots which will help us to better understand chemical signatures in DLAs.
For example, halos with high stellar masses all have abundance ratios tightly distributed around [C/O]$\approx -0.45$. These halos have a large number of metal-enriched stars, and the IMF is statistically well sampled, which leads to convergence in abundance ratios. On the other hand, for low-mass galaxies without Pop III stars (created through external enrichment), there is a larger scatter in [C/O] due to smaller numbers of stars leading to variations from stochastic sampling of the IMF. These galaxies are the points distributed along a vertical line on the left-hand side of their respective panels in Figure \ref{fig:scatter}. The maximum abundance value of ${\rm [C/O]}\sim 0.6$ is obtained for gas purely ejected by Pop III stars. Interestingly, this is roughly the maximum value found in the DLA observations. 
The max [C/O] value is generally achieved via one of two pathways which both occur in low-stellar-mass systems. In the first, the galaxy is purely Pop III. In the second path, there is a mixture of Pop III and metal-enriched stars, but the cumulative mass of metal-enriched stars is sufficiently low that when sampling the IMF no metal-enriching supernovae were created.

We can also see how varying our semi-analytic model parameters impacts the distribution of [C/O] in halos. Increasing the Pop III star formation efficiency (to $f_{\rm III}=0.01$) makes a substantial difference compared to the fiducial model. In Figure \ref{fig:scatter}, we see that it increases both the fraction of stellar mass in Pop III stars and [C/O], moving the distribution of points up and to the right. In Figure \ref{fig:scatter2}, we see that for systems with low stellar mass, the increase in Pop III star formation boosts [C/O] values. However, in galaxies with more than $\sim 10^6~M_\odot$ in stars, Pop III mass will be a negligible fraction in any of our parameterizations. 
 We note that decreasing the metal-enriched star formation efficiency, $f_{\rm II}$, would have a similar effect as increasing $f_{\rm III}$. However, we regard $f_{\rm III}$ as much more uncertain since $f_{\rm II}$ is calibrated with observations of high-redshift galaxies \citep{2020ApJ...897...95V}. 
We also note that in our model we assume that after the initial burst of Pop III star formation only metal-enriched star formation occurs. In the simulations of \cite{2022ApJ...935..174S} a subgrid prescription is utilized to follow inhomogeneous metal mixing in the ISM. This leads to a Pop III SFRD roughly an order of magnitude higher than our fiducial model. Thus, roughly speaking incorporating this effect could potentially be similar to our increased $f_{\rm III}$ case. However, if the additional Pop III due to inhomogeneous mixing takes a substantial amount of time to occur, it may be diluted by metal-enriched stars and have a limited impact on metal abundance ratios.

Another model variation which substantially changes the appearance of the scatter plots is using the older Pop III critical mass from \cite{2020ApJ...897...95V}. Because the values of $M_{\rm crit}$ in this case are higher, all star formation is more easily suppressed in low-mass halos which reduces the density of points in Figures \ref{fig:scatter} and \ref{fig:scatter2} (though the overall trends remain similar to the fiducial model). 

Additionally we see that increasing $t_{\rm delay}$ leads to significant changes in the [C/O] distribution. We note that increasing the delay time not only increases the duration when a halo contains purely Pop III stars, it also decreases the amount of metal-enriched stars that form once the delay is over. This is because the Pop III stars in our model rapidly ionize the IGM in their immediate surroundings. This leads to an exponential suppression in metal-enriched star formation in halos below $M_{\rm ion}$ (meant to capture the impact of photoheating of the ionized gas) given by a factor of $\exp(-t_{\rm rei}/[0.1t_{\rm H}])$, where  $t_{\rm rei}$ is the time elapsed since reionization occurred (different for each halo) and $t_{\rm H}$ is the age of the universe when reionization occurred. Increased $t_{\rm delay}$ then leads to less metal-enriched star formation in low-mass halos due to an increased $t_{\rm rei}$. We see the impact of this effect in Figure \ref{fig:scatter}, where halos are spread to higher Pop III fractions and higher [C/O] for $t_{\rm delay}=5\times 10^7$ yr. Additionally, in Figure \ref{fig:scatter2} halos are spread to lower total stellar mass in cases where the Pop III fraction and [C/O] are  high as a result of the ionization feedback effect just described. As discussed below, this parameter variation agrees best with the observational data due to the increased abundance of high-[C/O] DLAs.

Finally, we turn to comparing the predicted DLA [C/O] abundances to the observations of \cite{Sodini2024}. We assume the same distribution of QSO redshifts as in \cite{Sodini2024}, and make predictions for the expected number of DLAs observed above $z=5.9$ (our N-body simulations do not extend to lower redshifts). In order to match the total number of DLAs, we increase $A_{\rm eff}$ given by Eq.~\ref{eqn:A_eff} by a constant factor of $\approx 4$. As mentioned above, this may be reasonable since the values are extrapolated from lower-redshift simulations.
We also examine one case where we have reduced $M_0$ in $A_{\rm eff}$ by a factor of 10. In this case, lower mass halos contribute more strongly to the DLAs and instead of a correction factor of $\approx 4$ to the normalization of  $A_{\rm eff}$, the correction factor is ${\sim}0.3$.

\begin{figure}
\centering 
\includegraphics[width=12 cm]{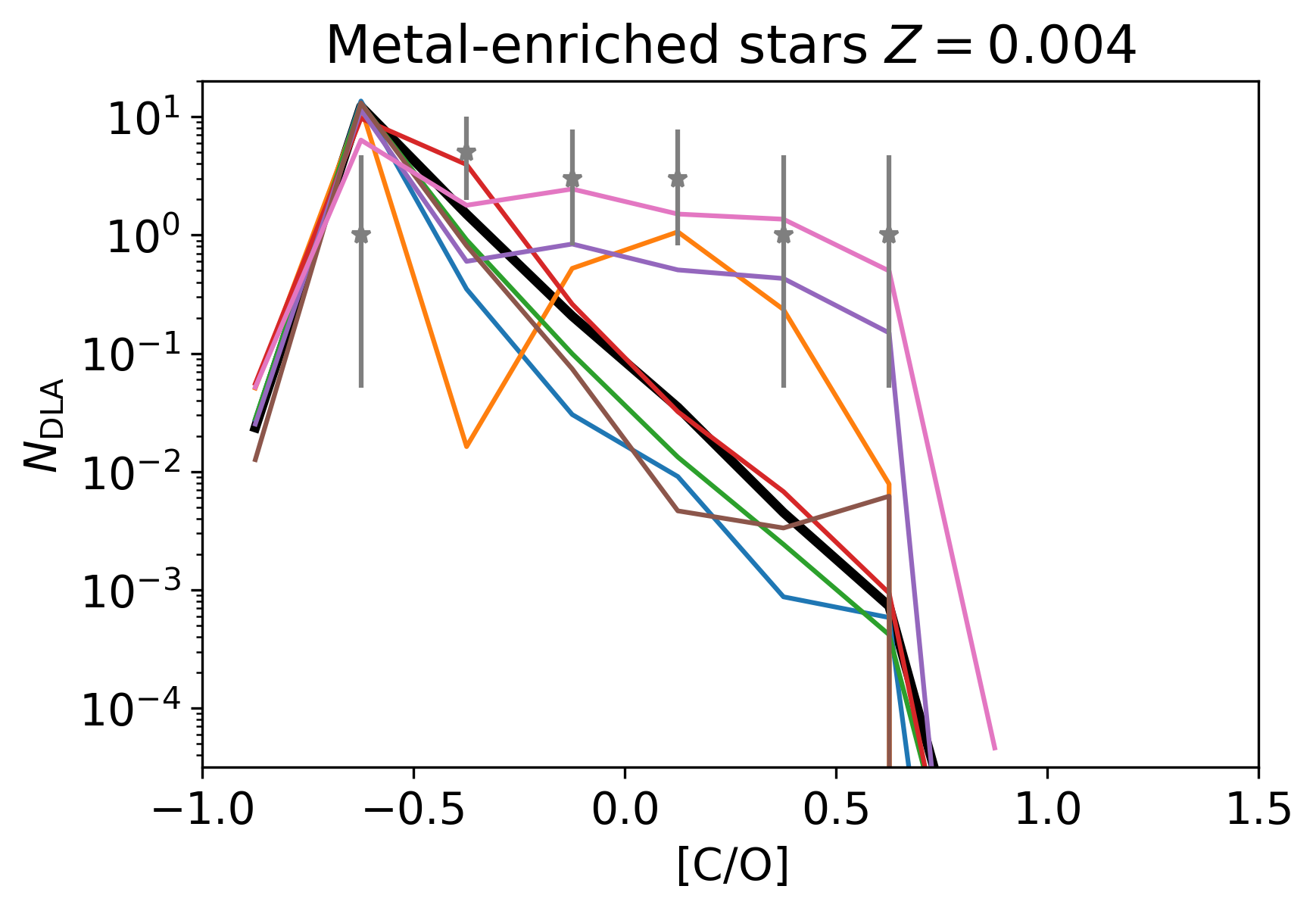}
\includegraphics[width=12 cm]{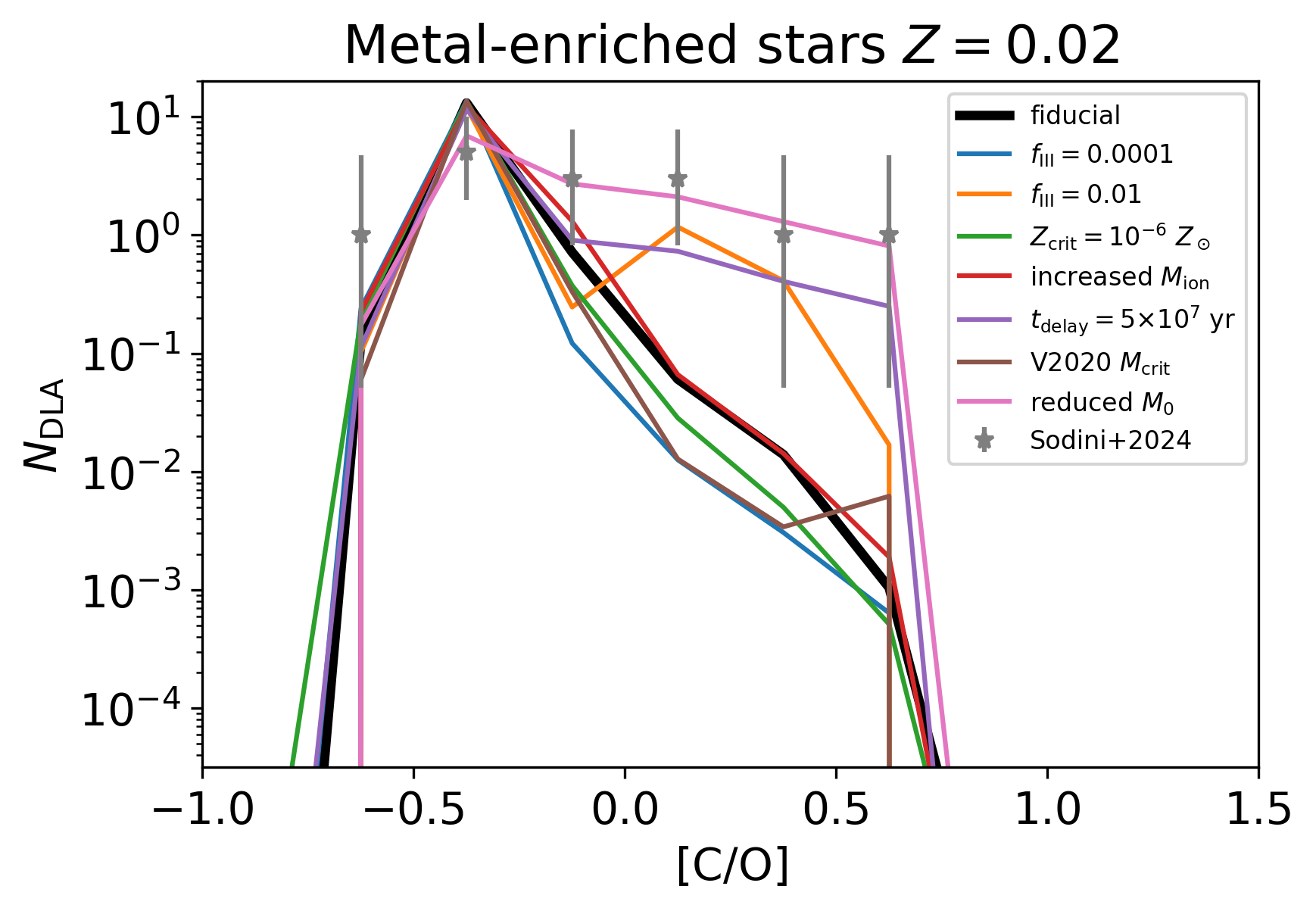}
\caption{\label{fig:N_DLA} Number of DLA analogs in each [C/O] bin as a function of [C/O] observed in \cite{Sodini2024} compared with our model predictions (for the same set of background QSOs). Here we only consider QSO spectra and DLA analogs above $z=5.9$ because our simulations do not extend below these redshifts. The observational data from \cite{Sodini2024} are compared with the model variations described in Figure \ref{fig:SFRDs}. We have added the ``reduced $M_0$'' model, which is the same as the $t_{\rm delay}=5\times 10^7$ yr case but with $M_0$ reduced by a factor of ten compared to the value in Eq.~\ref{eqn:A_eff}. We also present two cases where the metal-enriched supernovae yields are assumed to come from different metallicity stars (using the results from \cite{2006ApJ...653.1145K}).
Note that we modify the effective DLA cross section, $A_{\rm eff}$, by a constant factor for each model parameterization such that the total DLA abundance is consistent with the value inferred from the observed data. For most models, this involves increasing  $A_{\rm eff}$ in Eq.~\ref{eqn:A_eff} by ${\approx}4$. The only exception is the ``reduced $M_0$'' model where in addition to the $M_0$ reduction, $A_{\rm eff}$ is reduced by a factor of $\sim 0.3$.
The error bars on the observational points were computed as follows. Each bottom error bar extends to the mean value of $N_{\rm DLA}$ for a Poisson distribution that would reach or exceed the observational data point 5 percent of the time. Similarly, the upper error bars correspond to the mean value that would extend down to the observational point or lower 5 percent of the time.  }
\end{figure}

We present our predicted DLA abundances in Figure \ref{fig:N_DLA}, together with the corresponding observations from \cite{Sodini2024}.
We assume two different values of the metallicity of metal-enriched stars, which should bracket this uncertain parameter. We leave a fully self-consistent calculation where the precise metallicities of all stars are tracked for future work.
We also note that we do not include measurement errors in our predicted [C/O] values, but that the errors reported in  \cite{Sodini2024} are typically smaller than 0.1 and our bins are 0.25 in width. Thus, we would not expect this to have a substantial qualitative impact on our results.
The factor of $\approx 4$ correction mentioned above (which is essentially the same for all of the parameter variations, except for being $\approx0.3$ in the reduced $M_0$ case) does not change the shape of the theoretical curves in Figure \ref{fig:N_DLA}, only their normalization.

We see that in general, massive metal-enriched halos lead to a bump in DLA abundance at [C/O]$=-0.45(-0.6)$ for assumed metal-enriched stellar metallicity of $Z=0.02(0.004)$. Higher values of [C/O] require DLAs with larger contributions from Pop III stellar winds or small metal-enriched galaxies with more stochastic sampling of the IMF (though these cannot reach the highest observed [C/O] values).   Thus, one can infer the number of Pop III dominated systems expected in our different model parameterizations by examining the points in Figure \ref{fig:N_DLA} near [C/O]$\sim 0.6$. This number does not depend on our choice of Pop III metal yields.
Most of our models greatly underestimate the abundance of high-[C/O] DLAs compared to the observed data (3 orders of magnitude for our fiducial model).
However, two of our model parameterizations (increased $t_{\rm delay}$ and $f_{\rm III}$) predict that high-[C/O] DLAs containing Pop III signatures are expected to be found in the sample of \cite{Sodini2024}. The increased $t_{\rm delay}$ case in particular looks qualitatively similar to the observed data and agrees with the high-[C/O] DLA abundance to within the 95 percent confidence intervals defined in the caption of Figure \ref{fig:N_DLA} (as opposed to many orders of magnitude discrepancy in other models). This is an intuitive result since increased $t_{\rm delay}$ leads to larger ratios of Pop III to metal-enriched stars. We also examine one case with higher $t_{\rm delay}$ and $M_0$ in $A_{\rm eff}$ an order of magnitude lower than our fiducial model (which increases the importance of low-mass halos). This results in excellent agreement with the observations as discussed more in the following section.

\section{Summary and Conclusions}
We utilized a semi-analytic model of the first stars and galaxies (originally developed in \cite{2020ApJ...897...95V}) to make predictions for Pop III chemical signatures in high-redshift ($z\sim 6$) DLAs. This was motivated by recent observations of OI DLA analogs from \cite{Sodini2024} with chemical abundance ratios that are potentially indicative of Pop III enrichment \cite{Vanni2024}. Our modeling of Pop III DLAs is the first to include important physical effects such as 3D treatments of LW feedback, reionization, and IGM metal enrichment.

Following \cite{Kulkarni2013}, we adopted an effective area of our DLAs, $A_{\rm eff}(M_{\rm h})$, based on extrapolating the simulations of \cite{2008MNRAS.390.1349P}. This leads to a majority of DLAs being hosted by halos with virial masses between $M_{\rm h}=10^{8-9.5}~M_\odot$. Larger halos do not contribute substantially because of their low number densities, and smaller halos do not contribute substantially because of their low effective areas. We find rough agreement with the overall number of DLA analogs found in \cite{Sodini2024}, but require $A_{\rm eff}(M_{\rm h})$ to be increased by a factor of ${\sim}4$. This is unsurprising given that we are extrapolating simulation results from lower redshifts ($z=3$) in order to calibrate the effective area. In Figure \ref{fig:N_DLA}, we also show one case where in addition to a normalization change (of $\sim 0.3$ instead of $\sim 4$), $A_{\rm eff}(M_{\rm h})$ is modified by reducing $M_0$ by a factor of ten. This increases the contribution from low-mass halos and leads to excellent agreement with the observational data.

We focus our analysis on the [C/O] abundance ratio. For our assumed metal yields and IMFs, halos with [C/O]$>0$ generally have a non-negligible metal contribution from Pop III stars (see Figure \ref{fig:scatter}). We predict the number of DLAs observed from a sample of quasar spectra from $z=5.9-6.5$ as a function of [C/O]. These results demonstrate that DLAs are promising for studying Pop III stars in two ways.
First, our calculations show that, for reasonable parameter choices, one expects to find DLAs that contain the chemical signature of Pop III stars (which here corresponds to essentially any positive values of [C/O]). Second, changes in our semi-analytic model parameters dramatically impact the predicted abundance of high-[C/O] DLAs, which suggests that these data have the potential to put strong constraints on highly uncertain parameters related to Pop III star formation.  

We find that our fiducial model is strongly ruled out when compared with the observational data from \cite{Sodini2024} (see Figure \ref{fig:N_DLA}). It produces several orders of magnitude fewer DLAs with [C/O]$>0.5$ than the observations. Additionally, we individually vary other model parameters from their fiducial values and find that most of these parameterizations also underpredict the abundance of high [C/O] DLAs. However, increasing the delay time between Pop III and metal-enriched star formation to $t_{\rm delay}=5\times10^7~{\rm yr}$ leads to relatively close agreement with the data (within the 95 percent confidence intervals defined in the caption of Figure \ref{fig:N_DLA}). When increasing the delay time by the same amount and decreasing $M_0$ in $A_{\rm eff}$, there is excellent agreement with the data.
We note that this value of the delay time is reasonable as it corresponds to roughly half of the dynamical time at the virial radius of halos at $z\sim 6$. It is also in the range of values found from the simulations of \cite{2014MNRAS.444.3288J} (though we note those simulations are at much higher redshift and in lower mass halos).
We do not expect to achieve perfect agreement with the observations both because we have not performed a comprehensive search of the model parameter space and because  of our simplifying assumptions (e.g., our extrapolated treatment of the DLA effective area and the assumption of perfect metal mixing in halos). 

We emphasize that our predictions rely on several necessary simplifications.
First, we assume that halos retain one half of the metals produced by their stars. 
For the metals that are retained, we further assume that they become uniformly mixed throughout the interstellar medium on short time scales. We also adopt a simplified prescription in which all Pop III stars have yields that match the winds of rapidly rotating $35~M_\odot$ stars with $v_{\rm rot}/v_{\rm crit} = 0.6$ \citep{2023MNRAS.526.4467J}. 
In addition, we assume that the effective cross section of DLA analogues is equal to the effective area found in the low--redshift simulations of \citep{2008MNRAS.390.1349P}. 
All of these assumptions should be tested and refined with future hydrodynamical cosmological simulations of early galaxy evolution.

We note that despite these simplifications, our model represents a significant improvement. 
The previous model of \cite{Kulkarni2013} does not include several key processes that are required to compute the Pop III abundance with accuracy. These processes include LW feedback, the baryon-dark matter streaming velocity, halo merger histories, and a 3D treatment of metal enrichment and reionization feedback in the intergalactic medium. 
As a result, we find substantially different evolution of the Pop III star formation rate density. 
For example, \cite{Kulkarni2013} find that Pop III star formation ends entirely at $z \sim 8$, while we find that it continues through the end of our simulations at $z \sim 6$.

Our work is also the first to compare the data of \cite{Sodini2024} with the theoretically predicted number of absorption systems as a function of [C/O]. We find that Pop III metals are required to explain the high-[C/O] absorption systems in our model. 
For reasonable astrophysical parameters, and with only small adjustments from the fiducial values adopted prior to the release of the new data, we obtain abundances of objects with mixed Pop III and Pop II enrichment (that is, systems with [C/O] between 0 and 0.6) that agree well with the observations.

In future work, it will be beneficial to analyze high-redshift hydrodynamical radiative transfer simulations to obtain a more accurate treatment of the DLA effective area. We note that since the observed absorptions systems are identified with OI, technically future work will need to consider the effective ``DLA-analog'' area using OI column density rather than HI. If this area is larger for smaller halos than we assumed here, it could potentially increase the abundance of high-[C/O] DLAs in our predictions (as in our decreased $M_0$ example). This could occur due to the different UV background halos are exposed to prior to reionization compared to the low-redshift simulations we calibrate to here (which could alter the amount of self-shielded hydrogen gas within galaxies). 

It will also be important to investigate inhomogeneous metal mixing inside halos. This could leave pockets of unmixed Pop III supernovae ejecta, which may push more DLAs from low to high [C/O], possibly leading to closer agreement between our model and observations. Ultimately, it will be necessary to simultaneously analyze all available elemental abundance ratios and to perform a complete search over the model parameter space. This includes varying the Pop III IMF and models for elemental yields, both of which have substantial uncertainties (e.g., yields for many nonrotating Pop III models from \cite{2010ApJ...724..341H} do not have [C/O]$>0$). 

We also note that our 3 Mpc boxes are too small to properly model the topology of reionization \cite{2014MNRAS.439..725I}. Our reionization history is similar to the fiducial model shown in Figure 5 of \cite{2020ApJ...897...95V}, which realistically completes reionization at $z \sim 6$. However, given that the halo masses where Pop III star formation occurs depend strongly on whether each halo is reionizated or not, the Pop III abundance at the redshifts relevant here are expected to be sensitive to the precise history and topology of reionization. In future work we intend to couple our small-scale semi-analytic model to larger boxes ($>$100 Mpc across) simulated with {\sc 21cmFAST} \cite{2011MNRAS.411..955M}.

Finally, we emphasize the importance of extending these predictions to lower redshifts to understand exactly when Pop III chemical signatures are phased out, which should not depend on elemental yields from Pop III stars. 
Overall, the high sensitivity of our DLA abundance predictions to model parameters such as  the Pop III star formation efficiency and $t_{\rm delay}$ suggests that refined theoretical predictions will provide important constraints on the properties of Pop III stars and the smallest galaxies.

\acknowledgments
EV acknowledges the support of NSF grant AST-2009309 and NASA ATP grant 80NSSC22K0629. ZH and GB acknowledge support from NSF grant AST-2006176. All runs of the semi-analytic model were performed at the Ohio Supercomputer Center. 

\bibliography{DLA_project}
\end{document}